\documentclass[11pt,preprint]{aastex}
\usepackage{graphicx}

\begin{document}

\title{THE MILLENNIUM ARECIBO 21-CM ABSORPTION LINE SURVEY. III. \\
TECHNIQUES FOR SPECTRAL POLARIZATION AND RESULTS FOR STOKES $V$ \\
\today
}

\author{Carl Heiles}
\affil{Astronomy Department, University of California,
    Berkeley, CA 94720-3411; cheiles@astron.berkeley.edu}

\author{T.H. Troland}
\affil{Department of Physics and Astronomy, University of Kentucky,
Lexington, KY; troland@pa.uky.edu}

\begin{abstract}

	We outline the theory and practice of measuring the four Stokes
parameters of spectral lines in emission/absorption observations. We
apply these concepts to our Arecibo HI absorption line data and present
the results. We include a detailed discussion of instrumental effects
arising from polarized beam structure and its interaction with the
spatially extended emission line structure. At Arecibo, linear
polarization (Stokes $(Q,U)$ has much larger instrumental effects than
circular (Stokes $V$). We show how to reduce the instrumental
contributions to $V$ and to evaluate upper limits to its remaining
instrumental errors by using the $(Q,U)$ results. These efforts work
well for opacity spectra but not for emission spectra. Arecibo's large
central blockage exacerbates these effects, particularly for emission
profiles, and other telescopes with weaker sidelobes are not as
susceptible. We present graphical results for 41 sources; we analyze these
absorption spectra in terms of Gaussian components, which number 136,
and present physical parameters including magnetic field for each.

\end{abstract}

\tableofcontents

\section{INTRODUCTION}

	In February 1999 we used the Arecibo\footnote{The Arecibo
Observatory is part of the National Astronomy and Ionosphere Center,
which is operated by Cornell University under a cooperative agreement
with the National Science Foundation.} telescope to begin a series of
Zeeman-splitting measurements of the 21-cm line in absorption against
continuum radio sources. Heiles \& Troland (2001a, 2001b; papers I and
II) reported on a by-product of this survey, namely the Stokes $I$ data
from which spin temperatures and other information were derived. The
present paper focuses on the technical aspects of processing the
polarized spectral data and evaluating the instrumental errors. We also 
present the derived magnetic fields for 41 sources, which have 136
Gaussian components.

	There has been much discussion of HI Zeeman splitting
measurements because the polarized sidelobes, interacting with the
angular structure of the HI emission, can produce instrumental effects.
Even in the emission-absorption measurements presented in this paper,
these effects can in principle be serious. Therefore, we discuss these
effects in considerable detail. The bottom line is that instrumental
effects are evaluated for each source independently and, generally, are
negligible for the opacity spectra. However, for the expected emission
spectra they are not negligible. 

	\S \ref{stokestheory} outlines the basic theoretical concepts
involving Stokes parameters and emission/absorption lines. \S
\ref{stokespractice} discusses the least squares fitting process
required to extract the expected emission profile and the absorption
spectrum by combining calibrated on- and off-source data.  \S
\ref{generaldiscussion} discusses the physical reasons for and
contributors to instrumental effects that arise from polarized beam
structure. 

	The next three sections deal with instrumental contributions to
the {\it opacity} spectra. We begin by treating Stokes $V$ in  \S
\ref{problemevaluation}, discussing the empirical least-squares
evaluation and elimination of the two most basic instrumental effects,
namely the trigonometric dependences on parallactic angle $PA$ and $2PA$
(squint-like and squash-like dependences, respectively).  We use the
terms ``squint-like'' and ``squash-like'' for these
empirically-determined dependencies because they include contributions
from far-out sidelobes; in contrast, we use the terms ``squint'' and
``squash'' (sometimes preceded by the clarifier ``true'') for  the
contributions from only the primary beam and first sidelobe\footnote{The
terms squint and squash normally refer to only to the main beam, not
including the first sidelobe}. We then do the same in \S
\ref{empiricalqu} for the linearly polarized Stokes parameters $(Q,U)$.
We reach the important conclusion that instrumental effects in linear
polarization are about ten times larger than in circular. This allows us
to make an independent estimate of instrumental effects for $V$ using
$(Q,U)$, as we discuss explicitly in \S \ref{usingqu}.

	Our final discussion of instrumental effects (\S
\ref{emissionone}) addresses the reliability of Zeeman splitting results
for {\it emission} profiles. Here we independently evaluate ``true''
squint and squashand also the squint-like and
squash-like contributions. The difference between these is the
contributions from the far-out sidelobes. The far-out sidelobe
contribution is large for the Arecibo telescope. This
makes the instrumental effects quite serious for emission profiles.
Accordingly, we do not discuss the Zeeman splitting results for the
emission profiles. 

	Finally, \S \ref{results} presents the profiles for all sources
and a tabular list of the results for the Gaussian components. We select
a good sample for statistical analyses, successfully compare with previous
literature, and point out yet another source of uncertainty in the
derived magnetic fields.

\section{STOKES PARAMETERS OF SPECTRAL LINES IN  EMISSION/ABSORPTION}

\label{stokestheory}

\subsection{The ON and OFF spectra} 

\label{onoffspectra}

	Consider a particular polarization, which we designate by the
subscript $p$.  For the two circulars we have $p = LCP$ or $p
= RCP$, while for the linears we have {\it p = position angle}.  In the
presence of a continuum source that provides antenna temperature
$T_{src,p}$, the on-source antenna temperature is

\begin{equation} \label{one}
T_{src,p}(\nu) = T_{exp,p}(\nu) + T_{src,p} e^{-\tau _p(\nu)} \ ,
\end{equation}

\noindent where $T_{exp,p}(\nu)$ is the ``expected profile'', which is
the emission that would be observed in the absence of the source, and
$\tau _p(\nu)$ is the 21-cm line opacity, which depends on polarization;
both of these are functions of frequency because of the spectral line.
The appended symbol $(\nu)$ indicates frequency-dependent quantities
within the profile; unappended temperatures are continuum.

	We form Stokes parameters for the on-source antenna temperature
from arithmetic combinations of orthogonal polarizations $(p,p\!\!\!
\perp)$.  We designate the Stokes parameters $I_{src}, Q_{src}, U_{src},
V_{src}$ with the general symbol $S_{src,i}$, with $i = 0 \rightarrow
3$, respectively; the subscript $src$ designates on-source ``antenna''
Stokes parameters, derived from on-source antenna temperatures.  This
gives for Stokes $I_{src}$

\begin{mathletters}
\begin{equation}
\label{eqns0a}
S_{src,0}(\nu) = [T_{exp,p}(\nu) + T_{exp,p\perp}(\nu)] + 
[T_{src,p} e^{-\tau _p(\nu)} + 
T_{src,p \perp} e^{-\tau _{p \perp}(\nu)}] \ ,
\end{equation}

\noindent and for Stokes $(Q_{src}, U_{src}, V_{src})$

\begin{equation}
\label{eqns1a}
S_{src,i}(\nu) = [T_{exp,p}(\nu) - T_{exp,p \perp}(\nu)] + 
[T_{src,p} e^{-\tau _p(\nu)} - T_{src,p \perp} e^{-\tau _{p \perp}(\nu)}] \ , \ \
          (i = 1 \rightarrow 3) \ .
\end{equation}
\end{mathletters}

\noindent Here and below, $(i)$ implies $i = 1 \rightarrow 3$ unless
otherwise noted, and $p$ must correspond correctly with $i$. 

	We define 

\begin{mathletters} \label{taueqn1}
\begin{equation} 
\tau_0(\nu) \equiv { \tau_p(\nu) + \tau_{p_\perp}(\nu) \over 2}
\end{equation}

\noindent and we assume that the spectral line exhibits small
polarization, i.e.

\begin{equation}
	\tau_i(\nu) \equiv \tau_p(\nu) - \tau_{p \perp}(\nu) \ll 1  \ ,
\label{taudef}
\end{equation}
\end{mathletters}

\noindent where again $p$ must correspond correctly with $i$. Then we
expand equations~\ref{eqns0a} and \ref{eqns1a} and retain only the
lowest order terms in $\tau_i$, which are zeroth order for
equation~\ref{eqns0a} and first order for equation~\ref{eqns1a}.  This
gives

\begin{mathletters}
\begin{equation}
\label{eqns0b}
S_{src,0}(\nu) = [T_{exp,p}(\nu) + T_{exp,p \perp}(\nu)] + 
[T_{src,p} + T_{src,p_\perp}] e^{-\tau_0(\nu) }  \ ,
\end{equation}
\begin{equation}
\label{eqns1b}
S_{src,i}(\nu) = [T_{exp,p}(\nu) - T_{exp,p \perp}(\nu)] -
\tau_i(\nu) {[ T_{src,p} + T_{src,p \perp}] \over 2} e^{-\tau_0(\nu) } +
[ T_{src,p} - T_{src,p \perp}] e^{-\tau_0(\nu) }	 \ .
\end{equation}
\end{mathletters}

\noindent It is clearer to write the above expressing the temperature
sums and differences in terms of their Stokes parameters, which are the
appropriate sums and differences of antenna temperatures:

\begin{mathletters}
\label{eqns0}
\begin{equation}
\label{eqns0c}
S_{src,0}(\nu) = S_{exp,0}(\nu) + 
S_{src,0} e^{-\tau_0(\nu) } \ ,
\end{equation}
\begin{equation}
\label{eqns1c}
S_{src,i}(\nu) = S_{exp,i}(\nu) -
\tau_i(\nu) {S_{src,0} \over 2} e^{-\tau_0(\nu) } +
S_{src,i} e^{-\tau_0(\nu)} 	 \ .
\end{equation}
\end{mathletters}

\noindent Again, quantities subscripted with $exp$, like
$S_{exp,i}(\nu)$, are the frequency-dependent expected profile, i.e.\
what is expected to be observed at the source position if its continuum
flux were zero; quantities subscripted with $src$, like $S_{src,0}$,
are frequency-independent properties of the continuum source.

	Equation~\ref{eqns1c} for the polarized Stokes parameters $S_i$
consists of three terms: \begin{enumerate}

\item  The first term is the polarization of the expected emission
profile.  

\item The second term represents the polarized portion of the optical
depth $\tau_i(\nu)$ multiplying the attenuated Stokes $I_{src}$ (or
$S_{src,0}$) antenna temperature of the continuum source.  For Zeeman
splitting, this is the quantity of interest!

\item The third term represents the ordinary line opacity operating on
the polarized Stokes $Q,U,V$ (or $S_i$) antenna temperature of the
continuum source. 

\end{enumerate}

\subsection{ The ON-OFF spectra}

\label{onminusoff}

	For Stokes $I$, consider equation~\ref{eqns0c} and assume, for
the moment, that the spatial derivatives of $S_{exp,0}(\nu)$ are zero.
Then the two unknowns $S_{exp,0}(\nu)$ and $e^{-\tau_0(\nu) }$ are
easily separated observationally by taking on-source and off-source
measurements, for which $S_{src,0}$ changes from zero to the full source
intensity. More generally the spatial derivatives are nonzero; moreover,
Arecibo has significant sidelobes and we can never go completely off the
source. We account for these and other details by writing more
complicated versions of equation~\ref{eqns0c} and subject them to least
squares analyses. This is discussed in detail in \S 2 of paper I. 

	Now consider the polarized Stokes parameters in
equation~\ref{eqns1c} and assume that the spatial derivatives of their
expected emission profiles are zero. Then the combination of the second
and third terms is easily obtained by subtracting the on-source and
off-source measurements (ON--OFF). Below, as we did for $I$, we will
account for details using least square fits to more complicated
equations.  The third term in equation \ref{eqns1c} is of little
intrinsic interest because it reveals no new information: it is simply
the line opacity operating on the polarized flux. This term is often
large for Stokes $Q$ and $U$ because radio continuum sources often
exhibit significant linear polarization. 

	In contrast, the second term is of vital importance. In circular
polarization it is nonzero because of Zeeman splitting. In linear
polarization it reveals a correlation between the spatial structures of
the HI and the radio source Stokes parameters. Specifically, the more
highly polarized parts of the source produce larger fractional
contributions to the polarized opacity profile.  Unfortunately,
instrumental effects also contribute to the second term; we discuss
these in \S \ref{vproblems}. 

\subsection{ The particular case of Zeeman splitting}

\label{zmn}

	 Zeeman splitting is characterized by a frequency difference
between the two circular polarizations, so are concerned with the
subscript $i=3$. From now on we will write the subscripts with the
conventional notation $(Q,U,V)$ instead of $(1,2,3)$. We have

\begin{equation}
\tau_V(\nu) = {d \tau_0(\nu) \over d\nu} \delta \nu_Z
\end{equation}

\noindent As is well-known, for the 21-cm line $\delta \nu_Z = 2.8
B_{||}$ Hz, where $B_{||}$ is the line-of-sight field strength in
$\mu$Gauss.  To focus on the opacity spectrum, we consider only the
source terms in equation \ref{eqns0} (i.e., we set $S_{exp,0}(\nu)=0$)
and we assume no continuum circular polarization (i.e., we set
$V_{src}=0$). Writing the usual $I$ for $S_0$ and $V$ for $S_3$, we have

\begin{mathletters} \label{eqnzmn}
\begin{equation} \label{eqnzmn0}
I_{src}(\nu) = I_{src} e^{-\tau_0(\nu)}
\end{equation}
\begin{equation} \label{eqnzmn1}
V_{src}(\nu) = - {I_{src} \over 2}  e^{-\tau_0(\nu)} \tau_V(\nu)
\end{equation}
\end{mathletters}

\noindent or, in the most observationally relevant form,

\begin{equation}
\label{eqnzmn2}
V_{src}(\nu) = {d(I_{src}(\nu)/2) \over d\nu} \delta \nu_Z \ .
\end{equation}

\noindent This is exactly the same equation that applies to the
optically thin emission case.

        We recount this simple derivation to elucidate any possible
confusion about the role of Stokes $I$ opacity in deriving $\delta
\nu_Z$. This opacity weakens the $V$ spectrum; in particular, the effect
of the opacity difference $\tau_V(\nu)$ is weakened by the factor
$\exp(-\tau_0(\nu))$, and one might have expected this weakening to
reduce the derived value of $\delta \nu_Z$. This derivation shows that
using equation~\ref{eqnzmn2} provides the correct values of $\delta
\nu_Z$ under any circumstances, emission or absorption.

\section{EXTRACTING POLARIZED PROFILES FROM CALIBRATED SPECTRAL DATA}

\label{stokespractice}

\subsection{Stokes $I$} \label{stokespracticei}

	Paper I discussed our observing technique and the least-squares
fit for the Stokes $I$ expected profile, its spatial derivatives, and
the opacity profile. We observed a series of $N$ ``patterns'', denoted
by subscript $n$. Each pattern consists of a series of $J$ measurements
(subscript $j$), one being on source and the others being off source
displaced in different directions. This allows us to determine spatial
derivatives, which was important for the analysis of the Stokes $I$
profiles done in Paper II. Here we are concerned with the polarized
Stokes parameters, which are themselves detectable only with low
signal/noise and for which the spatial derivatives are expected to be
undetectable. Therefore, to begin our discussion we rewrite equation 8
in Paper I without the spatial derivative terms, obtaining a slight
generalization of equation \ref{one} above. 

\begin{equation} \label{ls1}
T_{ant,n,j}(\nu) = [T_{exp}(\nu)] +
[e^{-\tau (\nu)}] T_{ant,n,j} \ .
\end{equation}

\noindent $T_{ant,n,j}$, without the appended symbol $(\nu)$, is the
excess continuum antenna temperature over cold sky, which is usually
nonzero even for off source measurements because (1) telescope sidelobe
respond to the source, (2) the off position can lie within the primary
beam, and (3) diffuse continuum emission that happens to lie in the
source direction also contributes. $T_{ant,n,j}(\nu)$ includes the
effects of the HI line, while $T_{ant,n,j}$ does not---it is only the
continuum contribution.  $T_{ant,n,j}(\nu)$ is the antenna temperature,
i.e. the input to the receiver. 

	This equation applies to antenna temperatures measured in a
particular polarization. Therefore, it also applies to sums and
differences of antenna temperatures in orthogonal polarizations,
i.e.~the Stokes parameters.  For Stokes $I$ the equation barely changes.
However, the polarized Stokes parameters are slightly more involved. 

\subsection{Stokes $V$} \label{stokespracticev}

	First we treat the simpler case of Stokes $V$. When we form $V$
by subtracting RCP from LCP, the left hand side of equation~\ref{ls1}
becomes the measured value $V_{ant,n,j}(\nu)$.  In general, both the
source and $\tau(\nu)$ are polarized, so the equation becomes

\begin{mathletters} \label{eqnten}
\begin{equation} \label{vfit1}
V_{ant,n,j}(\nu) - V_{ant,n,j} e^{-\tau_0(\nu)} =
[V_{exp}(\nu)]   - \left [\tau_V(\nu) e^{-\tau_0(\nu)} \over 2 \right] 
I_{ant,n,j}  + [\Delta V_{n,j}(\nu)]
\end{equation}

\noindent where, as in equation \ref{eqns1c}, we retain only first-order
terms. Also, we have added an instrumental contribution $[\Delta
V_{n,j}(\nu )]$, which we discuss below.  Again, the square brackets
indicate quantities to be solved for by least squares. In a least
squares analysis we need on one side of the equation all of the
unknowns, and none of the knowns, which is why we transferred the
quantity  $V_{ant,n,j} e^{-\tau_0(\nu)}$ to the left-hand side; both
factors are known reasonably accurately from the observations. To make
the equations more concise, which is convenient for later discussion,
we make two definitions. First, we define the quantity

\begin{equation} \label{tauprime}
[\tau_V'(\nu)] \equiv - \left [\tau_V(\nu) e^{-\tau_0(\nu)} \over 2 \right] 
\end{equation}

\noindent which is the fractional circular polarization of the source's
absorbed flux ($V(\nu) \over I_{src}$ in equation \ref{eqnzmn1}), useful
because it is proportional to the frequency derivative of $I(\nu)$.
Second, we define $V'_{sky,n,j}(\nu)$ to be the first two terms on the
right hand side of equation \ref{vfit1}, i.e.

\begin{equation} \label{vfit2}
[V'_{sky,n,j}(\nu)] = [V_{exp}(\nu)] 
+ [\tau'_V(\nu)] I_{ant,n,j}
\ .
\end{equation}

\noindent so that equation \ref{vfit1} becomes

\begin{equation} \label{vfit3}
V_{ant,n,j}(\nu) - V_{ant,n,j} e^{-\tau_0(\nu)} =
  [[V'_{sky,n,j}(\nu)]]  + [\Delta V_{n,j}(\nu)]
\end{equation}
\end{mathletters}

\noindent where the double brackets around $[[V'_{sky,n,j}(\nu)]]$ serve
as a reminder that this term contains more than one unknown quantity.
The left hand side contains the measured quantities: $V_{ant,n,j}(\nu)$
is the channel-by-channel $V$ spectrum, while $V_{ant,n,j}$ (no $\nu$
dependence) is the continuum value, obtained from the off-line channels.
Similarly, $I_{ant,n,j}$ on the right hand side is the Stokes $I$
continuum value, also from the off-line channels. 

	For each spectral independently, the quantities in square
brackets are straightforwardly solved by least squares, except for the
instrumental contributions $[\Delta V_{n,j}(\nu)]$ for which the word
``straightforwardly'' does not necessarily apply. We either ignore this
contribution and estimate its magnitude, as discussed in \S
\ref{twopointfive}, or assume a functional dependence on parallactic
angle and include this in least squares fit (\S
\ref{problemevaluation}).

\subsection{Stokes $Q$ and $U$}

\label{stokespracticequ}

	A discussion similar to that of \S \ref{stokespracticev} applies
to Stokes $Q$ and $U$, but it becomes more complicated because the
measured values depend on parallactic angle $PA$, which changes with
hour angle. That is, including the instrumental error terms we have

\begin{mathletters}
\begin{eqnarray}
\label{skymatrix1}
\left[ 
\begin{array}{c}
 Q_{ant, n, j}(\nu)  \\
 U_{ant, n, j}(\nu)   \\ \end{array} \right] = 
{\bf R_n \ \cdot} 
\left[
\begin{array}{c}
 {Q_{sky, n,j}(\nu)}   \\
 {U_{sky, n,j}(\nu)}   \\ 
\end{array} \right] + 
\left[
\begin{array}{c}
 { \Delta Q_{n, j}(\nu) }  \\
 { \Delta U_{n, j}(\nu) } \\ 
\end{array} 
\right] \; .
\end{eqnarray}

\noindent where

\begin{eqnarray}
\label{skymatrix4}
{\bf R_n} = 
\left[
\begin{array}{cc}
  \cos 2PA_{n} & \sin 2PA_{n} \\
 -\sin 2PA_{n} & \cos 2PA_{n} \\
\end{array} \; \right] 
\end{eqnarray} \ .
\end{mathletters}

\noindent Here we neglect the change in $PA$ during a pattern, so the
terms are subscripted only with $n$. As in the above discussion for $V$,
for the purpose of the least squares fit we must retain all of the
unknowns, and none of the knowns, on the right hand side of the
equation. For the least squares fit, the required equation is the
analogy to equation \ref{vfit3}

\begin{eqnarray}
\label{skymatrix6}
\left[ 
\begin{array}{c}
 Q_{ant, n, j}(\nu)  \\
 U_{ant, n, j}(\nu)   \\ 
\end{array} \right] - 
{\bf R_n \ \cdot} 
\left[
\begin{array}{c}
e^{-\tau_0(\nu)} Q_{sky, n, j}  \\ 
e^{-\tau_0(\nu)} U_{sky, n, j}  \\
\end{array} \right] =
{\bf R_n \ \cdot} 
\left[ \left[
\begin{array}{c}
Q_{sky, n, j}'(\nu) \\ 
U_{sky, n, j}'(\nu) \\ 
\end{array} \right] \right] +
\left[
\begin{array}{c}
 {[ \Delta Q_{n, j}(\nu) ]}  \\
 {[ \Delta U_{n, j}(\nu)  ]} \\ 
\end{array} 
\right] \; .
\end{eqnarray}

\noindent We remind the reader that $Q_{sky, n, j}$ (no $\nu$
dependence) is the continuum value obtained from the off-line channels,
and is known quite accurately.  The quantity $Q_{sky}'(\nu)$, with the
prime, is defined analogously to that for $V_{sky}'(\nu)$ in equation
\ref{vfit2}. As with $V$, the quantities (not the matrices) in square
brackets are solved by least squares. We either ignore the instrumental
contributions and estimate their magnitudes, as discussed in \S
\ref{twopointfive},  or include their dependences on $PA$ in the least
squares fit (\S \ref{problemevaluationqu}).

	For the continuum values of $(Q,U)$, the major contributor to
$(\Delta Q, \Delta U)$ is zero offsets. We measure two polarized Stokes
parameters, Stokes $U$ and $V$, by crosscorrelating the voltages of
orthogonal polarizations and the third, Stokes $Q$, by differencing the
powers of orthogonal polarizations (Heiles 2001). The last is
particularly susceptible to instrumental problems, primarily a
frequency-independent zero offset, both because the receiver
temperatures differ and because the gains of the two polarizations are
not perfectly calibrated. The shape of the spectrum can also be changed
by the introduction of a weak replica of the Stokes $I$ spectrum, but we
ignore this because it is indeterminate. Even the crosscorrelation
spectra have small offsets because of instrumental coupling between the
two channels, and these depend slightly on the Stokes $I$ value because
of errors in the Mueller matrix coefficients (see Heiles et al 2001b).

	These offset errors are frequency independent, so they affect
only the continuum values. $\Delta U_{n,j}$, which is determined by
crosscorrelation, is smaller and more nearly constant than $\Delta
Q_{n,j}$. In equations \ref{vfit3} and \ref{skymatrix6}, the quantities
enclosed in square brackets are unknown and need to be determined by
least squares fitting. The number of unknowns is awkwardly large, as was
the case for equation 9 in Paper I, and for the same reasons. We apply
the same iterative technique here, namely neglecting the $n$-dependence
of $Q_{sky, j}(\nu)$ and $U_{sky, j}(\nu)$ and first solving for the set
of $J$ values for each spectral channel individually, and then solving
for the $NJ$ values of $\Delta Q_{n, j}$ and $\Delta U_{n, j}$ using all
channels and measurements simultaneously.  Having done this, we correct
the measured $Q_{ant,n,j}(\nu), U_{ant,n,j}(\nu)$ values by subtracting
the frequency-independent offsets $\Delta Q_{n, j}$ and $\Delta U_{n,
j}$, and use those corrected values to proceed with the least squares
solution of equation \ref{skymatrix6}, including frequency-dependent
instrumental contributions as discussed in \S \ref{problemevaluation}.

\section{ INSTRUMENTAL PROBLEMS WITH POLARIZED STOKES PARAMETERS:
GENERAL DISCUSSION}
\label{vproblems} 

\label{generaldiscussion}

	This section discusses the most important instrumental
contributions to instrumental polarization. Polarized opacities are
small, so we must consider systematic effects at very low levels.  The
most important contributors include the following: \begin{enumerate}

	\item There is instrumental coupling between the polarized
Stokes parameters and Stokes $I$, which creates replicas of the Stokes
$I$ line in the polarized Stokes parameter spectra. These couplings are
described by the Mueller matrix (Heiles et al 2001b) and are corrected
for, but the corrections are imperfect. 

	Experience teaches us that this is usually the dominant
instrumental contribution to the measured Stokes $V$ spectra, typically
amounting to a few tenths of a percent. Fortunately, in deriving Zeeman
splitting its presence is unimportant: observers normally call this a
``gain error'' and use a least squares technique to remove it. 

	\item The Mueller matrix is derived from observations of
small-diameter continuum sources, so it applies to beam center. 
However, it changes within the telescope beam. For Stokes $V$ the
primary effect is  ``beam squint'', for which the $V$ beam has positive
and negative lobes on opposite side of beam center. This arises from the
two polarizations pointing in slightly different directions. Beam squint
interacts with the first spatial derivative of the Stokes $I$ profile to
produce a false contribution. We assume that the beam squint is fixed
with respect to the feed, so its false contribution varies periodically
with the parallactic angle $PA$. 

	Beam squint is theoretically predicted to occur for Stokes $V$
but not for $Q$ and $U$. However, Arecibo has significant beam squint in
$Q$ and $U$ as measured by the response to the first spatial derivative
in Stokes $I$; it is probably produced by the significant aperture
blockage. Figures 12, 13, and 14 of Heiles et al (2001a) show that the
effective beam squint for both $Q$ and $U$ is about ten times that for
$V$.

	Beam squint correlates with the angular structure in the {\it
emission} line to produce unreal features in the {\it polarized} Stokes
emission and absorption spectra. Experience shows that for Stokes $V$,
beam squint interacting with the first derivative of HI intensity is the
most serious instrumental problem (Heiles 1996). For conventional
telescopes this contribution can be measured and removed rather
accurately. For the Arecibo telescope, however, the polarized beam
pattern changes partly systematically, partly erratically with hour
angle, making the correction impractical.

	We can estimate the magnitude of this contribution from the
measured angular derivatives of the Stokes $I$ expected profile and the
approximately-known magnitude of the beam squint (\S
\ref{twopointfive}). Also, we can determine the approximate contribution
by fitting for the change in the apparent polarized opacity spectrum as
the polarized beam rotates with respect to the sky with hour angle
(\S\ref{problemevaluation}). However, because Arecibo's beam
characteristics, including squint, change while tracking, the squint
contribution varies not only with $PA$ but also has a semi-random
component.  We least-square fit equation  \ref{trigeq} for the
systematic squint component, as discussed below, but the semi-random
component remains.

	\item For Stokes $Q$ and $U$, the primary effect for the
polarized beam is ``beam squash'', for which the polarized beams have a
four-lobed cloverleaf pattern, with two positive lobes on opposite side
of beam center and two negative ones rotated $90^\circ$. This arises
from the two polarizations having slightly different beamwidths. Beam
squash interacts with the second spatial derivative of the Stokes $I$
emission profile to produce a false contribution that varies
periodically with twice the parallactic angle $2PA$. 

	Beam squash is theoretically predicted to occur for Stokes $Q$
and $U$, but not for $V$. However Arecibo, with its significantly
blocked aperture, violates this rule. As with beam squint, Figures 12,
13, and 14 of Heiles et al (2001a) show that Arecibo's effective beam
squash for both $Q$ and $U$ is about ten times that for $V$.

	For  Stokes $V$, most telescopes have negligible beam squash, so
our experience with them does not apply to Arecibo. We can estimate the
magnitude of the squash contribution from the measured angular
derivatives of the Stokes $I$ expected profile and the
approximately-known magnitude of the beam squash (\S
\ref{twopointfive}). Also, we can approximately determine the
contribution by fitting for the change in the apparent polarized 
spectrum as the polarized beam rotates with respect to the sky with hour
angle (\S\ref{problemevaluation}); this works well for the opacity
spectrum. However, it does not work well for the expected emission
spectrum because of the incomplete $PA$ coverage (\S \ref{emissionone}).

	For Stokes $Q$ and $U$ we can estimate the squash contribution
from the measured spatial derivatives and the known beam squash.
However, we cannot determine it by fitting for the change with $2PA$
because real linear polarization also varies with $2PA$. 

	\item All four Stokes parameters are affected by ``far-out
sidelobes'', i.e.\ sidelobes outside the first sidelobe. These arise
primarily from ordinary diffraction and, secondarily, from surface
inaccuracies. Far out sidelobes for Arecibo are particularly strong
because of the severe aperture blockage. And they contribute
disproportionately to the squint and squash response. They are likely to
be polarized comparably to the first sidelobe, and the HI spatial
structure within them changes more because the angular differences are
larger; this magnifies their importance. The first sidelobe itself
serves as an illustration: for a uniform extended source in Stokes $I$,
it contributed 0.34 as much as the primary beam in
2000\footnote{Arecibo's surface was readjusted and the focus point
changed in 2002. The current fraction is much lower, $\lesssim 0.10$.});
yet in Stokes $V$ it contributed almost {\it twice} as much to the
observed squint response as does the primary beam! (see Table 1 and
Figure 14 of Heiles et al 2001a). 

	Far-out sidelobes can, like squint and squash, be expressed as a
Fourier series in $PA$. The first term is squint-like and the second is
squash-like. Below, we use the terms {\it squint-like} and {\it
squash-like} to denote the contribution from the telescope, i.e.\ from
the primary beam and all sidelobes, and unless the context dictates
otherwise we use the terms {\it squint} and {\it squash} to denote the
contributions from the primary beam and first sidelobe\footnote{To be
precise, true squint and squash refer only to the main beam without the
first sidelobe. But Heiles et al (2001a) measured the influence of
both---for Arecibo, the first sidelobe is large because of the large
aperture blockage. For this reason, in this paper we stretch the
definition of ``true''.}.

\end{enumerate}

	Contribution (1) produces only a replica of the Stokes $I$
opacity profile; in Zeeman splitting this is routinely removed by
least-squares fitting and has no damaging effect. Accordingly, we do not
consider it further.

	The  contributions (2), (3), and (4) exhibit sinusoidal
dependences on either $PA$ (squint-like) or $2PA$ (squash-like). There
are two independent ways to evaluate these contributions, each with its
own problem. The {\it empirical} way (\S \ref{problemevaluation}) is
observing a range of PA and performing a least squares fit on the
results to directly evaluate (and remove) the sinusoidal dependences;
the problem, especially for the emission profile, is the incomplete $PA$
coverage. The other way (\S \ref{twopointfive}) {\it predicts} the
instrumental effects using the angular derivatives of the HI structure
in the sky, together with the already known polarization structures of
the Arecibo beam; the problem is the neglect of the far-out sidelobe
contribution. We turn to a detailed discussion of these issues in the
next few sections.

\section{EMPIRICAL EVALUATION OF SQUINT-LIKE AND SQUASH-LIKE
CONTRIBUTIONS TO STOKES $V$ OPACITY SPECTRA} 

\label{problemevaluation}

	Squint-like structure in the Stokes $V$ polarized beam interacts
with the first spatial derivative of the Stokes $I$ profile to produce a
false contribution to $V$ that varies periodically with the parallactic
angle $PA$, and squash-like structure interacts with the second spatial
derivative to produce a contribution that varies periodically with
$2PA$. Here we follow Heiles (1996) in using least squares fitting to
empirically evaluate both contributions. If we had complete and uniform
$PA$ coverage these fits would be straightforward, but this is not the
case. 

	Beam squint-like and squash-like structure produce instrumental
contributions $\Delta V_{n,j}(\nu) = V_{sqnt,n,j}(\nu)$ and
$V_{sqsh,n,j}(\nu)$, respectively, where $\Delta V_{n,j}(\nu)$ is the
instrumental contribution in equation \ref{vfit3}. We parameterize these
as follows:

\begin{mathletters} \label{trigeq}
\begin{equation} \label{trigeq1}
V_{sqnt, n}(\nu) =  [V_{sqnt, cos}(\nu)] \cos(PA_n) + 
[V_{sqnt, sin}(\nu)] \sin(PA_n) 
\end{equation}
\begin{equation} \label{trigeq2}
V_{sqsh, n}(\nu) = [V_{sqsh, cos}(\nu)] \cos(2PA_n) + 
[V_{sqsh, sin}(\nu)] \sin(2PA_n) 
\end{equation}
\end{mathletters}

\noindent Again, the square brackets indicate the unknown quantities to
be determined by least squares.  Note that the continuum quantities do
not appear in this equation; this is because the squint and squash
contributions arise only from the HI {\it emission} in the vicinity of
the source (as long as the continuum source is small compared to the
beam), so the source intensity is irrelevant. Moreover, we are assuming
no spatial gradients in $V_{sqnt,cos}$ and $V_{sqnt,sin}$, so have
dropped the subscript $j$. 

	Least squares fitting equation \ref{eqnten} allows us to solve
for six unknowns, i.e.\ the two contained in $V'_{sky}(\nu)$ in equation
\ref{vfit2} together with the four instrumental ones in equations
\ref{trigeq}, as long as we cover a sufficiently large range in $PA$. At
Arecibo, the maximum $PA$ range is $\sim 180^\circ$; for other alt-az
telescopes, a significantly larger range is available only for sources
that are nearly circumpolar.  

\begin{figure}[h!]
\begin{center}
\includegraphics[width=3.5in] {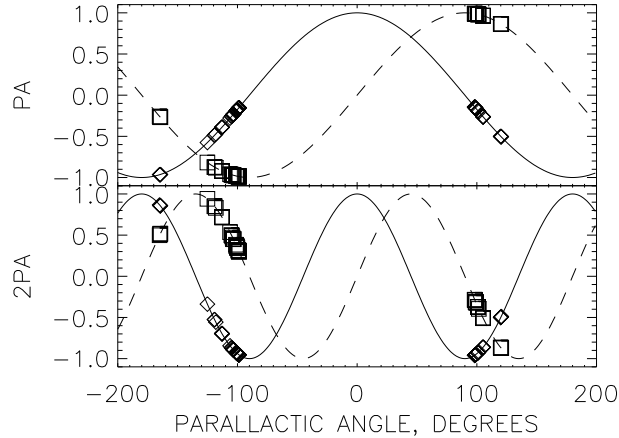}
\end{center}

\caption{Trigonometric functions (solid, cosine; dashed, sine) of $PA$
and $2PA$ versus $PA$. The squares and diamonds depict the actual
observed $PA$'s for 3C138 and illustrate the absence of complete coverage.
\label{3C138paplot} }
\end{figure}

	Figure \ref{3C138paplot} shows the $PA$ coverage for our example
of 3C138, which is unusual in having rather sparse $PA$ coverage. This
sparse coverage occurs for two reasons: first, the declination of 3C138
differs from Arecibo's latitude by only $\sim 2^\circ$, which means
that the $PA$ changes rapidly near transit. Thus, as the source is
observed and the $PA$ increases from $\sim 90^\circ$ to $\sim 270^\circ$
(same as $-90^\circ$), a broad band of $PA$ centered near $180^\circ$ is
not sampled. Second, the exigencies of scheduling meant that it received
little time at negative hour angles.  Most sources have more complete
coverage. We choose 3C138 as the example because its incomplete $PA$
coverage highlights the associated difficulties and because the angular
derivatives of brightness temperature are unusually high, which
exacerbates instrumental effects.

	The incomplete $PA$ coverage has different ramifications for the
derived expected and opacity profiles: \begin{enumerate}

	\item For the expected emission spectrum $V_{exp}(\nu)$ the
effects of incomplete $PA$ coverage are serious. Five terms in equation
\ref{vfit1} are independent of being on or off source. These are the
five emission-profile terms, namely $V_{exp}(\nu)$ in equations
\ref{eqnten} and the four terms in equation \ref{trigeq}. These five terms
constitute the first three frequency components of a sin/cosine Fourier
series with real coefficients.  If the $PA$ coverage were complete and
uniform, covering the {\it full} range of $360^\circ$ with uniform
sampling, then the least-squares solution of equation \ref{vfit1} would
be identical to a regularly-sampled  Fourier transform in which the five
terms are orthogonal and therefore independent.  The incomplete coverage
of $\lesssim 180^\circ$ is a problem because it removes this
orthogonality. The nonorthogonality  produces coupling between all five
terms and, also, extra noise. We will refer to this problem in the
ensuing discussion under the rubric ``covariance''.

	\item For the opacity spectrum $\tau'_{V}(\nu)$ the effects of
incomplete $PA$ coverage are much less serious.  The reason is simply
that $\tau'_{V}(\nu)$ is, fundamentally, the ON-OFF spectrum and the
continuum source is strong: the $PA$ coverage enters only peripherally.
There is little covariance between $\tau'_{V}(\nu)$ and the other five
terms discussed above. 

\end{enumerate}

\subsection{ Stokes $V$: General Discussion}

\label{problemevaluationv}

	We have our choice in what to include for $\Delta V_{n,j}(\nu)$
in the least squares fits to equation \ref{eqnten}.  We could fit by
ignoring both the squint-like and squash-like contributions, i.e.\
setting $\Delta V_{n,j}(\nu)=0$; we denote these solutions for the 2
unknowns $V_{exp}(\nu)$ and $\tau'_V(\nu)$ in equation \ref{vfit3} with
the additional subscript 0.  We could include only the squint-like terms
in $V_{sqnt}(\nu)$, for which we have 4 unknowns (the original 2 plus
$[V_{sqnt, cos}(\nu)]$ and $[V_{sqnt, sin}(\nu)])$, for which the
additional subscript is 1; or we could include both squint-like and
squash-like terms, for which we have 6 unknowns and the additional
subscript is 2. 

	The instrumental contributions in these three cases are most
concisely expressed by matrix equations that give the contribution to
the two derived quantities [$V_{exp}(\nu)$ and $\tau'_V(\nu)$] in
equations \ref{eqnten} in terms of the two true ones and, in addition,
the four true squint and squash ones in equation \ref{trigeq}. Here, by
``true'' we mean the values that actually occur, which are not equal to
the ones derived by the least squares fit because of the covariance
caused by incomplete $PA$ coverage. We calculate these matrix elements
by inserting known artificial signals into the data and calculating the
resulting contributions. For example, to evaluate the effect of
$V_{sqsh,cos, true}$, we insert the artificial signal $\cos (2PA_{n,j})$
into spectral channel number 2 (which contains no useful astronomical
information) and process it identically to the other spectral channels;
the resulting values of $\tau'_V$, $V_{exp}$, etc., provide the
corresponding matrix elements. The matrix elements are the approximate
coupling coefficients between a given derived quantity, such as
$\tau'_V$, and the squint-like and squash-like instrumental effects as
embodied in, for example, $V_{sqsh,cos,true}$. These coefficients differ
for each source because of the differing $PA$ coverage. In these fits, we
assume $\tau_0(\nu)=0$; the matrix elements depend somewhat on
$\tau_0(\nu)$, so the particular values shown below are only
representative.

For the three cases we have:

\subsubsection{Least-squares fitting including neither squint nor squash}

	First, we derive $\tau'_{V,0}(\nu)$ and $V_{exp,0}(\nu)$ by
not fitting for squint and squash, i.e.\ by least squares fitting
equation \ref{vfit3} setting the instrumental $\Delta V_{n,j}(\nu)=0$. 
For 3C138, this gives

{\footnotesize
\begin{eqnarray}
\label{trigeqn0138}
\left[ 
\begin{array}{c}
\tau'_{V, 0} \\
V_{exp, 0} \\
\end{array} \right] =
\end{eqnarray}
$$
\left[
\begin{array}{cccccc}
  1.01 & 6.8\textrm {e-5 K}^{-1} & -6.7 \textrm {e-5 K}^{-1} & -1.1 \textrm {e-5 K}^{-1} & -2.1 \textrm {e-5 K}^{-1} & -3.3 \textrm {e-5 K}^{-1} \\
  +3.8 \textrm { K}  &  0.98   &   -0.29  &  -0.21  & -0.74    &  +0.12   \\
\end{array} \right] \cdot 
\left[
\begin{array}{c}
\tau'_{V, true} \\
V_{exp, true} \\
V_{sqnt, cos, true} \\
V_{sqnt, sin, true} \\
V_{sqsh, cos, true} \\
V_{sqsh, sin, true} \\
\end{array} \right]  \; .
$$
}

	First we discuss $\tau_{V,0}'$, whose matrix elements are in the
first row. The first matrix element, 1.01, reflects the fact that our
least-squares solution for $\tau_{V,0}'$ actually works and returns the
correct value. The second expresses the contribution of real Zeeman
splitting in the expected emission profile [$V_{exp}(\nu)$] to the
derived $\tau_{V,0}'$, and the remaining ones express the instrumental
(fake) contributions arising from the polarized beam interacting with
spatial derivatives in [$I(\nu)$]. The numerical values for these last
five matrix elements have upper limits, which we obtain as follows. 
From equations \ref{eqnzmn1} and \ref{tauprime}, we find that the
opacity $\tau'_{V}(\nu) = {V(\nu) \over I_{src}}$. For 3C138, $I_{src}
\approx 116$ K, so in calculating $\tau'_{V,0}(\nu)$ the instrumental
contribution $V(\nu)$ gets multiplied by $\sim {1 \over 116} = 8.5
\times 10^{-3} \ {\rm K}^{-1}$. This is the upper limit for the last
five matrix elements and it is well above the actual values.

	In particular, all of the matrix elements on the first row
(except the first) are inversely proportional to the continuum source
intensity: the stronger the source, the weaker the instrumental
contribution to $\tau'_{V}$. This large reduction in instrumental
effects is one reason why Arecibo, with its huge collecting area, is the
instrument of choice for these single-dish absorption studies.

	Now we discuss $V_{exp,0}$, whose matrix elements are in
the second row. Again, the second matrix element, 0.98, reflects the
fact that our least-squares solution for $V_{exp}(\nu)$ returns the
correct value. Regarding the squint and squash terms, this least squares
calculation of $V_{exp,0}$ ignores all $PA$ variation and is therefore
equivalent to an average of all the measured values. The third matrix
element (which multiplies $V_{sqnt,cos,true}$) equals $-0.29$. This must
equal the average value of $\cos(PA)$ because any nonzero value of
$V_{sqnt,cos,true}$ contributes this fractional amount in a straight
average. Visual inspection of figure \ref{3C138paplot} confirms this
expectation. Similarly, the sine component $V_{sqnt,cos,true}$ equals
$-0.21$. This component has the possibility of being nearly zero if the
source has symmetric hour angle coverage, because $\sin(PA)$ is
antisymmetric; indeed, this is the case for many of our sources---but
not for 3C138 because of the exigencies of scheduling.

\subsubsection{Least-squares fitting including squint but not squash}

	Next, we least squares fit equation \ref{vfit3} including
only the two squint-like terms in equation \ref{trigeq} and ignoring
squash.  The matrix becomes

{\footnotesize
\begin{eqnarray}
\label{trigeqn1138}
\left[ 
\begin{array}{c}
\tau'_{V, 1} \\
V_{exp, 1} \\
V_{sqnt, cos, 1} \\
V_{sqnt, sin, 1} \\
\end{array} \right] = 
\end{eqnarray}
$$ \left[
\begin{array}{cccccc}
    1.01   &  +6.8 \textrm {e-5 K}^{-1} & -5.4 \textrm {e-5 K}^{-1} & -7.2 \textrm {e-5 K}^{-1} & -4.8 \textrm {e-5 K}^{-1} & -9.1 \textrm {e-6 K}^{-1} \\
   +4.7 \textrm { K} &    0.99           & 0                &   0              &   -1.33           &   -0.10          \\
   +3.1 \textrm { K} &   0.02            &  0.98            &      0           &  -2.02           &   -0.40          \\
   -0.2 \textrm { K} &   0               &  0               &  0.99            &     0            & -0.48            \\
\end{array} \right] \cdot 
\left[
\begin{array}{c}
\tau'_{V, true} \\
V_{exp, true} \\
V_{sqnt, cos, true} \\
V_{sqnt, sin, true} \\
V_{sqsh, cos, true} \\
V_{sqsh, sin, true} \\
\end{array} \right]  \; . $$
}

	First we discuss $\tau_{V,1}'$. Comparison of the third and
fourth elements in the top rows of equations \ref{trigeqn0138} and
\ref{trigeqn1138} shows that including the squint changes the
contributions to $\tau'_{V,1}$ by factors of 0.8 and 6.5 for the cosine
and sine components, respectively. The sine component contributes more
to $\tau_{V,1}'$ than to $\tau_{V,0}'$, which is surprising. As we noted
above, 3C138 has asymmetric $PA$ coverage; sources having more nearly
symmetric $PA$ coverage exhibit large reductions in the squint
contribution to $\tau'_{V,1}$. For example, 3C207 has nearly symmetric
coverage and the factors above are much smaller---0.3 and 0.02,
respectively.

	Now we discuss $V_{exp,1}$. Because we explicitly fit for
them here, the $V_{exp,1}(\nu)$ (second row) matrix elements of both
components of $V_{sqnt, true}$ (i.e.\ the third and fourth elements)
are essentially zero. However, there is a large covariance between
$\cos(PA)$ and $\cos(2PA)$ for the incompletely sampled $PA$ range,
which leads to the large value of $-1.33$ for the matrix element for
$V_{sqsh, cos, true}$. It is surprising that this element exceeds unity.
Similarly, the last two matrix elements for $V_{sqnt,cos,1}$ also exceed
unity. Such surprises can occur when fitting nonorthogonal functions
with high covariance (these functions being the incompletely sampled
trigonometric functions of $PA$ and $2PA$).

\subsubsection{Least-squares fitting including both squint and squash}

	Finally, we least squares fit equation \ref{vfit3} including
both the two squint-like and squash-like terms in equation \ref{trigeq}.
Then we get

{\footnotesize
\begin{eqnarray}
\label{trigeqn2138}
\left[ 
\begin{array}{c}
\tau'_{V, 2} \\
V_{exp, 2} \\
V_{sqnt, cos, 2} \\
V_{sqnt, sin, 2} \\
V_{sqsh, cos, 2} \\
V_{sqsh, sin, 2} \\
\end{array} \right] = 
\end{eqnarray}
$$
\left[
\begin{array}{cccccc}

    1.01     &  +6.8 \textrm {e-5 K}^{-1}       &   -5.4 \textrm {e-5 K}^{-1}   &   -7.2 \textrm {e-5 K}^{-1} &   -6.3 \textrm {e-5 K}^{-1} &    +1.1 \textrm {e-5 K}^{-1} \\
    +4.4 \textrm { K} & 1.02         & 0         &    +0.01  &    -0.05  & -0.01      \\
    +2.7 \textrm { K} &  +0.06       &   0.98    &    +0.01  &    -0.08  & -0.01      \\
      0    \textrm { K}   & +0.01   & 0         &     1.00  &    -0.01  & -0.01          \\
    -0.3  \textrm { K} &  +0.02       & 0        &        0  &     0.96  &    0      \\
    +0.3  \textrm { K}   & 0        &  0         &     +0.02 &    -0.01  &    0.98    \\
\end{array} \right] \cdot
\left[
\begin{array}{c}
\tau'_{V, true} \\
V_{exp, true} \\
V_{sqnt, cos, true} \\
V_{sqnt, sin, true} \\
V_{sqsh, cos, true} \\
V_{sqsh, sin, true} \\
\end{array} \right]  \; .$$
}

	For $\tau_{V,2}'$, including squash gives a modest degradation
for the two squash terms (the last two matrix elements in the top row). 
More impressively, for $V_{exp,2}$ the large squash contributions to
$V_{exp,1}$ in equation \ref{trigeqn1138} (the last two matrix elements
in the second row) are changed by factors of $\sim 0.04$ and 0.1. This 
desired elimination of the squint and squash contributions to
$V_{exp,2}$ comes at a heavy price. As we shall see in \S
\ref{emissionone} and Figure \ref{3C138pafigexp1.ps}, the spectrum
$V_{exp,2}$ is much noisier than for $V_{exp,0}$ and $V_{exp,1}$;
unfortunately, this means that we cannot provide reliable Stokes $V$
spectra for the emission spectra. The excess noise is again the
covariance caused by the incomplete $PA$ coverage. As is true with any
least squares fit, coefficients having high covariance are determined
with large errors.

	Comparison of the above matrices shows that we can significantly
improve our resistance to instrumental effects by observing as full and
complete $PA$ range as possible. Because most sources have better
complete $PA$ coverage than 3C138, their opacity profiles $\tau_{V,2}'$
have smaller instrumental contributions than either $\tau_{V,0}'$ or
$\tau_{V,1}'$. Accordingly, we shall include both squint-like and
squash-like terms in the fit to equation \ref{vfit3} and always present
$\tau_{V,2}'$. The derived quantity $\tau'_{V, 2}$ automatically has the
squint-like and squash-like contributions removed. 
	
\subsection{ Instrumental contributions to $\tau'_{V}(\nu)$ for the
example of 3C138}

\label{3C138examplev}

\subsubsection{Stokes V: the data}

	We illustrate these concepts by showing and discussing the
instrumental contributions to the $\tau'_V(\nu)$ spectra for 3C138. We
choose 3C138 because it is one of the few sources to exhibit a clearly
detectable signal in $\tau'_{V}(\nu)$ and because its $PA$ coverage is
not very good, so it should represent a less-than-optimum case.

\begin{figure}[h!]
\begin{center}
\includegraphics[height=6.75in]{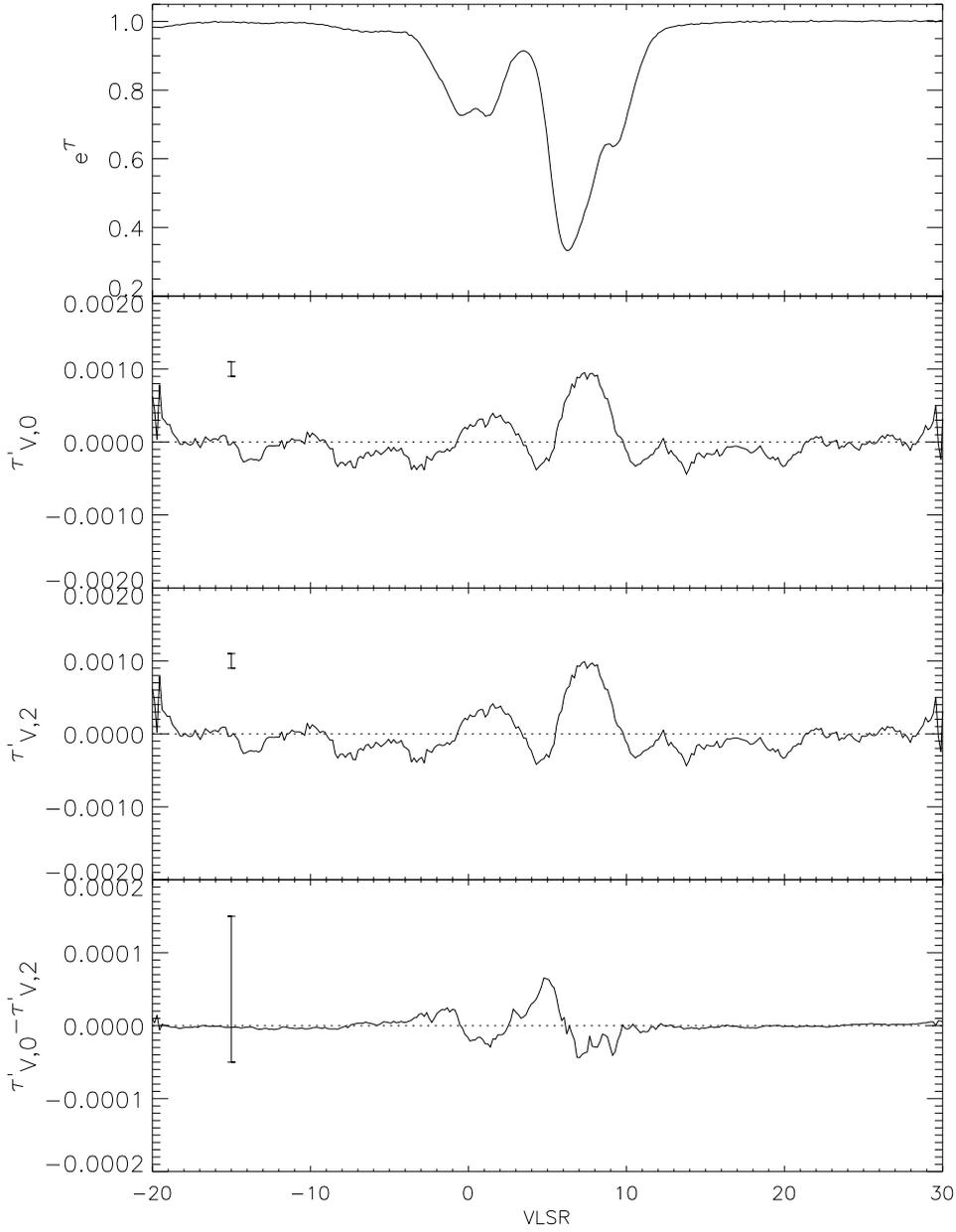}
\end{center}

\caption{$I$ and $V$ opacity spectra for 3C138.  The first (top) panel
is $e^{-\tau_0 (\nu)}$.  The second panel is $\tau'_{V,0}(\nu)$, derived
by ignoring squint and squash; the third is $\tau'_{V,2}(\nu)$, derived
including both squint and squash; and the fourth is the difference
$\tau'_{V,0}(\nu) - \tau'_{V,2}(\nu)$.  The spectra in the bottom three
panels are boxcar smoothed by 9 channels to reduce the noise. As shown
by the sample errorbar, the scale on the bottom panel is expanded by a
factor of ten. \label{3C138pafig1} } \end{figure}

	Figure \ref{3C138pafig1} illustrates three derived spectra for
$\tau'_{V}(\nu)$.  The first (top) panel is the classical opacity
spectrum $e^{-\tau_0 (\nu)}$.  The second panel is $\tau'_{V,0}(\nu)$,
which is derived not fitting for squint and squash, i.e.\ least squares
fitting equation \ref{vfit3} with $\Delta V_{n,j}=0$; this is equivalent
to the standard ON-OFF spectrum. The third panel is $\tau'_{V,2}(\nu)$,
which is derived including both squint-like and squash-like terms in
equation \ref{trigeq}. We don't show $\tau'_{V,1}(\nu)$ because it is
indiscernibly different from  $\tau'_{V,2}(\nu)$. Even
$\tau'_{V,0}(\nu)$ and $\tau'_{V,2}(\nu)$ in the second and third panels
are almost identical; the fourth panel shows the difference on a
ten-times expanded scale. The similarity of $\tau'_{V,0}(\nu)$ and
$\tau'_{V,2}(\nu)$  is a clear indication that neither squint- nor
squash-like effects contribute significantly to the Stokes $V$ opacity
profile for 3C138.

\subsubsection{ Stokes $V$: evaluation of matrix products for the
example of 3C138}

\label{stokesvsquintobs}

	We can evaluate the approximate instrumental contribution of the
$V(\nu)$ emission terms to $\tau'_{V,2}(\nu)$. These instrumental
contributions $\delta \tau'_{V}(\nu,2)$ are given by equation
\ref{trigeqn2138}, {\it viz.}

{\footnotesize
\begin{eqnarray}
\label{trigeqndiff138}
\left[ 
\begin{array}{c}
\delta \tau'_{V, 2} \\
\end{array} \right] = 
\end{eqnarray}
$$
\left[
\begin{array}{cccccc}
     0     &  +6.8 \textrm {e-5 K}^{-1}       &   -5.4 \textrm {e-5 K}^{-1}   &   -7.2 \textrm {e-5 K}^{-1} &   -6.3 \textrm {e-5 K}^{-1} &    +1.1 \textrm {e-5 K}^{-1} \\
\end{array} \right] \cdot
\left[
\begin{array}{c}
\tau'_{V, true} \\
V_{exp, true} \\
V_{sqnt, cos, true} \\
V_{sqnt, sin, true} \\
V_{sqsh, cos, true} \\
V_{sqsh, sin, true} \\
\end{array} \right]  \; .$$
}

\noindent From the lower two panels of figures \ref{3C138pafigcos} and
\ref{3C138pafigsin}, we crudely estimate

{\footnotesize 
\begin{eqnarray} \label{trigeqnfigs138} 
\left[ 
\begin{array}{c} \tau'_{V, true} \\ 
V_{exp, true} \\ V_{sqnt, cos, true} \\ 
V_{sqnt, sin, true} \\ V_{sqsh, cos, true} \\ 
V_{sqsh, sin, true} \\
\end{array} \right] 
\sim \left[  
\begin{array}{c} \dots \\ 
\dots \\ 
0.7 {\rm K} \\ 
0.1 {\rm K}  \\ 
0.2 {\rm K}  \\ 
0.2 {\rm K}  \\ 
\end{array}
\right]   \end{eqnarray}
}

\noindent These numerical estimates are zero-to-peak, not peak-to-peak. 

	We have not specified  the contribution from $V_{exp, true}$
because we cannot measure it accurately (see \S \ref{emissionone}).
However, we can estimate it. The Zeeman splitting in the emission line
is likely to be comparable to that in the absorption line. That is,
loosely speaking we expect ${V_{exp}(\nu) \over T_{exp}(\nu)} \sim
{\tau'_{V, 2}(\nu) \over \tau_0(\nu)}$. Roughly, if $\tau_0(\nu)
\lesssim 1$, then $T_{exp}(\nu) \approx T_s \tau_0(\nu)$, where $T_s$ is
the spin temperature. This gives 

\begin{equation} \label{vexpnu}
V_{exp}(\nu) \sim T_s \tau'_{V, 2}(\nu) 
\end{equation}

\noindent We obtain the contribution of $V_{exp}(\nu)$ to $\delta
\tau'_{V, 2}(\nu)$  by multiplying the above equation \ref{vexpnu} by
the corresponding matrix element (the second element in equation
\ref{trigeqndiff138}). This yields a contribution 

\begin{equation} \label{deltatau2}
{\delta \tau'_{V, 2}(\nu) \over\tau'_{V, 2}(\nu)} \sim 6.8 \times
10^{-5} T_s \lesssim 6.8 \times 10^{-3}   \ ,
\end{equation}

\noindent where we have assumed $T_s=100$ K for this estimate, which is
generously high given the results of Paper II. Therefore, the fractional
contribution ${\delta \tau'_{V, 2}(\nu) \over\tau'_{V, 2}(\nu)}$ is
negligible for any source flux and we can neglect the contribution of
$V_{exp}(\nu)$. 

	To get a rough estimate of the maximum total instrumental
contribution from squint- and squash-like effects, we add the absolute
values of the four individual contributions in equation
\ref{trigeqnfigs138}. We obtain $\delta \tau'_{V, 2} \lesssim 6 \times
10^{-5}$. This is comparable to the last panel in Figure
\ref{3C138pafig1}; both are about 20 times smaller than the detected
Zeeman splitting in 3C138. This is comfortably small. These instrumental
contributions scale inversely with source flux density $S$. For all
sources our plots (e.g.\ figure \ref{3C138zmndec02}) exhibit equation
\ref{trigeqn2138}'s matrix product for $\delta \tau'_{V, 2}(\nu)$. The
profile $\delta \tau'_{V, 2}(\nu)$ represents the approximate 
channel-by-channel instrumental contribution to the Stokes $V$ opacity
profile resulting from all squint- and squash-like contributions. In no
case is this contribution significant compared to $\tau'_{V, 2}(\nu)$.
This is fortunate, because it means that even if $\delta \tau'_{V,
2}(\nu)$ is not well-determined, subtracting its contribution incurs
little loss of accuracy in the final result $\tau'_{V, 2}(\nu)$.

\section{EMPIRICAL EVALUATION OF SQUINT-LIKE AND SQUASH-LIKE
CONTRIBUTIONS TO STOKES $(Q,U)$ OPACITY SPECTRA} 

\label{empiricalqu}

	In their discussion of polarized sidelobes, Heiles et al (2001a)
provide numerical coefficients for true squint and squash of Stokes
$(Q,U)$. These coefficients are about ten times larger than for Stokes
$V$. We believe that squint and squash are representative samples of all
polarized beam effects, so that this indicates that all sidelobes are
more serious in linear than circular polarization. Specifically, we
assume that this factor of ten applies not just to true squint and
squash, but also to squint-like, squash-like, and all other types of
sidelobe contribution.

	The polarized spectra $\tau'_{Q,1}(\nu)$ and $\tau'_{U,1}(\nu)$
should be zero unless there is opacity structure in the HI that varies
across the source together with continuum polarization that also varies
across the source. If we assume that there is no such structure, then
any nonzero behavior in $\tau'_{Q,1}(\nu)$ and $\tau'_{U,1}(\nu)$ must
result from the instrumental contribution of polarized sidelobes.
Dividing these by ten provides an estimate of the instrumental
contribution to Stokes $\tau'_{V,2}(\nu)$.

\subsection{ Stokes $(Q,U)$: General Discussion}

\label{3C138examplequ}

	Stokes $Q$ and $U$ are more complicated to treat than $V$
because the sky values are rotated as in equation \ref{skymatrix1}.
Moreover, we cannot fit for squash-like behavior because its
$PA$-dependence is identical to that of real linear polarization. After
performing the correction for the continuum offsets described in \S
\ref{stokespracticequ}, we are left with

\begin{eqnarray}
\label{skymatrix5}
\left[
\begin{array}{c}
 {[ \Delta Q_{n}(\nu) ]}  \\
 {[ \Delta U_{n}(\nu)  ]} \\ 
\end{array} 
\right] = 
\left[
\begin{array}{c}
 {[Q_{sqnt,cos}(\nu)]} \cos PA_n +  {[Q_{sqnt,sin}(\nu)]} \sin PA_{n} \\
 {[U_{sqnt,cos}(\nu)]} \cos PA_n +  {[U_{sqnt,sin}(\nu)]} \sin PA_{n} \\
\end{array} 
\right] \; .
\end{eqnarray}

\noindent where, as with $V$ in equation \ref{trigeq}, we have dropped
the $j$ subscript because we assume spatial derivatives are zero.  

	We could derive matrix elements for Stokes $(Q,U)$ as we did for
$V$ in equations \ref{trigeqn0138} and \ref{trigeqn1138} (We cannot
derive those for squash-like behavior).  However, we will not do this.
The matrix elements depend on the $PA$ coverage and Stokes $I$; the only
difference between linear and circular polarization is the necessity to
include the additional $PA$ dependencies in equations \ref{skymatrix1}
and \ref{skymatrix5}, and the matrix elements for $(Q,U)$ would be
comparable in magnitude to those for $V$ but different in detail. Owing
to the illustrative nature of our discussion, it is not worth taking up
this space. 

\subsection{ Examples of observed linear polarization
$[\tau'_{Q}(\nu),\tau'_{U}(\nu)]$ }

\label{problemevaluationqu}

\subsubsection{The example of 3C138}

\begin{figure}[h!] \begin{center}
\includegraphics[height=6.5in]{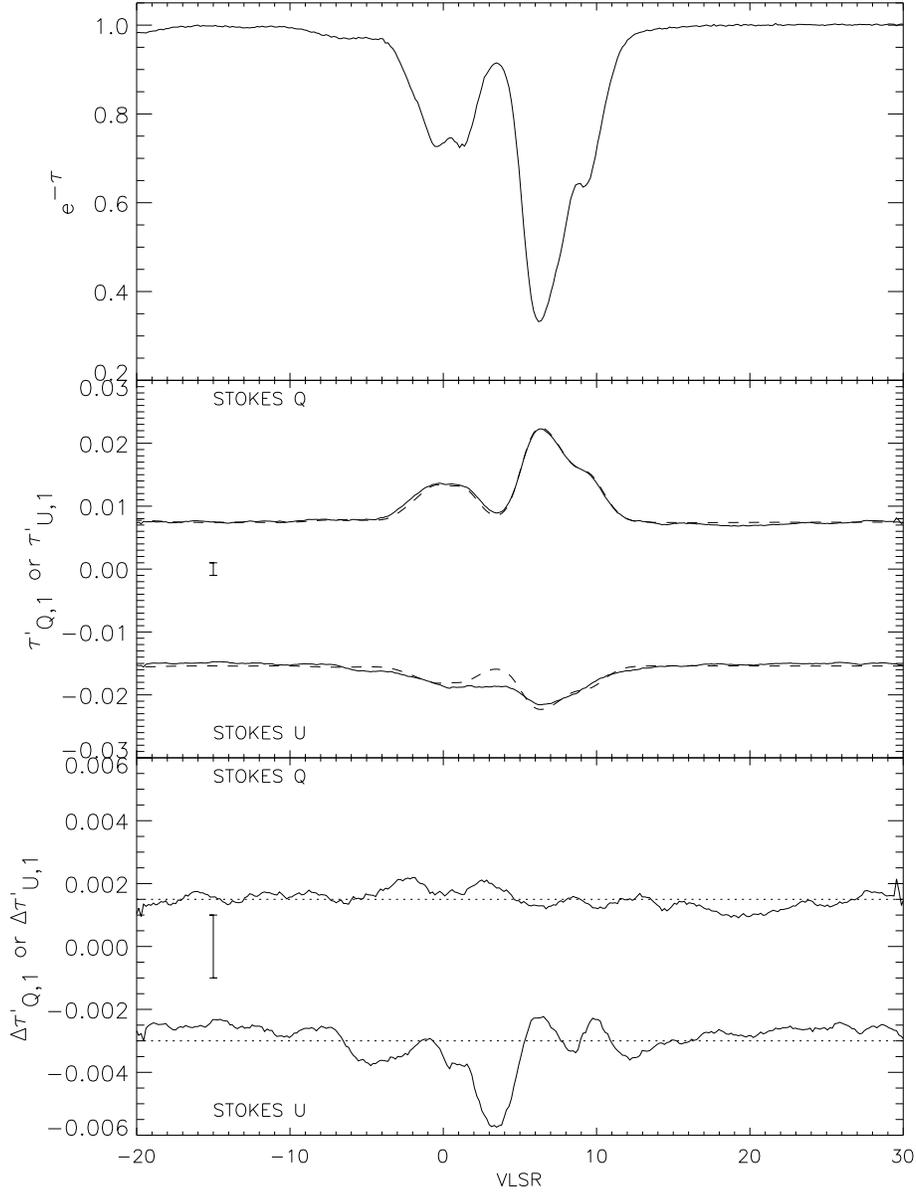} 
\end{center}

\caption{$I$ and $(Q,U)$ spectra for 3C138. The first (top) panel is
$e^{-\tau_0 (\nu)}$.  The second panel shows $\tau'_{Q,1}(\nu)$ and
$\tau'_{U,1}(\nu)$, derived including squint and ignoring squash. 
The dashed lines are the fits of
the $e^{-\tau_0 (\nu)}$ spectra to the $\tau'(\nu)$ spectra. The
bottom panel shows the ``gain-corrected'' versions of $\tau'_{Q,1}(\nu)$
and $\tau'_{U,1}(\nu)$, equal to the difference between the solid and
dashed lines in the second panel. Spectra in the bottom two panels have
displaced zeros and are boxcar smoothed by 9 channels for clarity. 
\label{3C138pafigqu00} } \end{figure}

	Figure \ref{3C138pafigqu00} illustrates the derived spectra for
$\tau'_{Q}(\nu)$ and $\tau'_{U}(\nu)$. The first (top) panel shows the
classical opacity spectrum $e^{-\tau_0 (\nu)}$.  The second (middle)
panel shows $\tau'_{Q,1}(\nu)$ and $\tau'_{U,1}(\nu)$, which are derived
including the squint, i.e.\ least squares fitting equation
\ref{skymatrix6} with $(\Delta Q_{n,j}, \Delta U_{n,j})$ given by
equation \ref{skymatrix5}. 

	The two spectra in the middle panel show features with peak
excursion $\sim 0.015$. However, their shapes mimic to some degree the
shape of the Stokes $I$ opacity profile $\tau_0(\nu)$ in the top panel.
This similarity in shape is probably the result of a small error in the
Mueller matrix coefficients, causing Stokes $I$ to leak into $Q$ and $U$
at the level of a few tenths of a percent. This is the gain error
discussed in item 1 of \S \ref{generaldiscussion}.

	The bottom panel of figure \ref{3C138pafigqu00} shows
gain-corrected spectra of $\tau'_{Q}(\nu)$ and $\tau'_{U}(\nu)$.  We fit
the two middle-panel spectra to $\tau_0(\nu)$ on a channel-by-channel
basis; the dashed lines in the middle panel are the fits. The bottom
panel shows the residuals [$\Delta \tau'_{Q,1}(\nu), \Delta
\tau'_{U,1}(\nu)]$, i.e.\ the data minus the fitted points. These
residual profiles are the results with the gain error removed, and
should be zero. They are, in fact, zero except for bumps at the $\sim
0.003$ level. 

 	Under our assumption of zero true polarization, these $\sim
0.003$ bumps must be the instrumental contribution from polarized
sidelobes. Moreover, this instrumental contribution must arise from
non-squint-like contributions because squint-like behavior has been
removed from $\tau'_{Q,1}(\nu)$ and $\Delta \tau'_{U,1}(\nu)$. There are
four possible production mechanisms for these bumps. The first is that
there really is true linear polarization, i.e.\ that the true
$\tau'_{Q,true}(\nu)$ and/or $\tau'_{U,true}(\nu)$ are not equal to
zero. We will first dispose of this possibility by considering a
different source, 3C454.3. 

\label{tlps138}

\subsubsection{The example of 3C454.3}

\label{tlps}

	3C454.3 is a particularly useful source for understanding
linearly polarized sidelobes because it is a VLBI calibrator and has a
very small angular size  $(\sim 14$ milliarcsec; Fomalont et al 2000;
this is about 1000 times smaller than 3C138). Despite the existence of
tiny scale atomic structure (reviewed by Heiles 1997), we expect
$\tau'_Q(\nu)$ and $\tau'_U(\nu)$ to be very small. We assume that any
departure from zero is an instrumental contribution from polarized
sidelobes. 

	Here we will need the squint matrix for 3C454.3, the equivalent
of equation \ref{trigeqn1138}, which is 

{\footnotesize
\begin{eqnarray} 
\label{trigeqn1} \left[  
\begin{array}{c}
\tau'_{V, 1} \\ 
V_{exp,  1} \\
V_{sqnt, cos, 1} \\ 
V_{sqnt, sin, 1} \\
\end{array} \right] =  
\end{eqnarray} $$ 
\left[ \begin{array}{cccccc}
1.01 & -1.9  \textrm {e-6 K}^{-1} & -1.7  \textrm {e-5 K}^{-1} &  -6.2  \textrm {e-6
K}^{-1} &  +5.5  \textrm {e-5 K}^{-1} &  +3.4  \textrm {e-5 K}^{-1} \\ 
1.3 \ {\rm K} & 1.01 & -0.01 & 0 & -1.42 & +0.14 \\
-5.4 \ {\rm K} & +0.04 & 0.97    &  +0.01  &  -2.12  & +0.22   \\ 
+1.1  \ {\rm K} & 0 & 0   &  1.00 &  +0.02 &  -0.67   \\
\end{array} \right] 
\cdot  \left[ \begin{array}{c} 
\tau'_{V, true} \\
V_{exp,  1} \\
V_{sqnt, cos, true} \\ V_{sqnt, sin, true} \\ V_{sqsh, cos, true} \\
V_{sqsh, sin, true} \\ \end{array} \right]  \; . $$ } 

\noindent Note, as anticipated in the discussion of equation
\ref{trigeqn0138}, that the matrix elements on the first row tend to be
inversely proportional to the source flux: 3C454.3 is about 2.4 times
more intense than 3C138. Also, of course, the above $4 \times 6$ matrix
elements for $V$ are comparable to those for $Q$ and $U$.

\begin{figure}[h!]
\begin{center}
\includegraphics[height=6.5in]{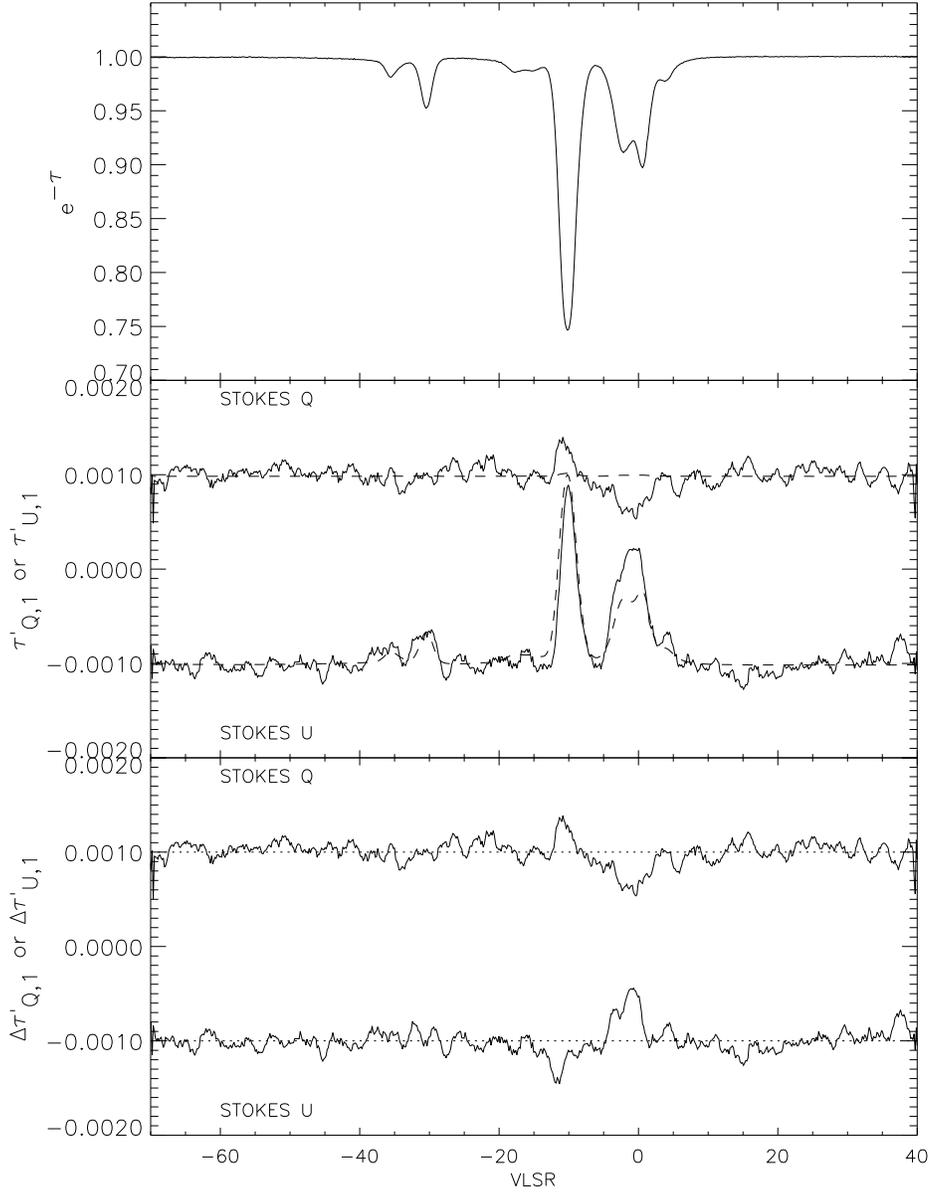}
\end{center} 

\caption{This figure is the 3C454.3 analog of figure
\ref{3C138pafigqu00}. The first (top) panel is $e^{-\tau_0 (\nu)}$.  The
second panel shows $\tau'_{Q,1}(\nu)$ and $\tau'_{U,1}(\nu)$, derived
including squint and ignoring squash. 
The dashed lines are the fits of
the $e^{-\tau_0 (\nu)}$ spectra to the $\tau'(\nu)$ spectra. The bottom panel shows the
``gain-corrected'' versions of $\tau'_{Q,1}(\nu)$ and
$\tau'_{U,1}(\nu)$, equal to the difference between the solid and dashed
lines in the second panel. Spectra in the bottom two panels have
displaced zeros and are boxcar smoothed by 9 channels for clarity.
\label{3C454.3pafigqu} } \end{figure}

	Figure \ref{3C454.3pafigqu} is the 3C454.3 equivalent of 3C138
figure \ref{3C138pafigqu00}, in which the middle and bottom panels show
the uncorrected ($\tau'_{1}(\nu)$) and gain-corrected ($\Delta
\tau'_{1}(\nu)$) spectra, respectively.  The spectra in the bottom panel
should be zero. They are, in fact, zero except for bumps at the $5
\times 10^{-4}$ level in both $\Delta \tau'_{Q,1}(\nu)$ and $\Delta
\tau'_{U,1}(\nu)$ near $VLSR=-12$ and $0$ km s$^{-1}$. These bumps must
be the instrumental contribution from polarized sidelobes. 

	The polarized sidelobes produce this contribution according to
the matrix elements of equation \ref{trigeqn1}, which for the
squint-like terms are about $1 \times 10^{-5}$ K$^{-1}$. A bump of $5
\times 10^{-4}$ in $\Delta \tau'_{U,1}(\nu)$ would need a combination of
the cosine and sine components of either $Q_{sqnt}$ or $U_{sqnt}$ to be
$\sim {5 \times 10^{-4} \over 1 \times 10^{-5} \ {\rm K}^{-1}} = 50$ K.
The actual values, not plotted here to save space and forestall the
wrath of the Almighty, are $\sim 0.4$ K. The bumps in $\Delta
\tau'_{1}(\nu)$ cannot be from squint-like behavior.

	Thus, for 3C454.3 $\Delta \tau'_{Q,1}(\nu)$ and $\Delta
\tau'_{U,1}(\nu)$ are nonzero while the true values should be zero. The
3C454.3 bumps are about 3 times smaller than the 3C138 ones, while
3C454.3 has flux 2.4 times larger; thus the ratios of the bump to the
source flux are comparable. This is roughly what's expected if the bumps
are caused by $V_{sqnt}$ and $V_{sqsh}$, because the first-row matrix
elements are roughly in the ratio of the source fluxes.  We conclude
that for both sources the production mechanism involves the polarized
beam interacting with angular derivatives of HI emission. Moreover,
there is no squint-like behavior in this interaction, because it has
been fitted for and thereby automatically subtracted out.

\subsection{Possible production mechanisms for fake linear polarization}

\label{linpolmech}

	The unreal Stokes $(\Delta \tau_{Q,1}(\nu), \Delta
\tau_{U,1}(\nu))$ bumps we see in 3C454.3 (and, probably, the bumps we
see in 3C138) cannot be from squint-like contributions, because these
have been removed in the least-squares fit. There are three possible
production mechanism for these unreal bumps: \begin{enumerate}

	\item One is the semi-random components of the squint-like
contributions $(Q_{sqnt}, U_{sqnt})$ (see item 2, \S
\ref{generaldiscussion}). From Heiles et al (2001a), these are smaller
than the uniform components of $(Q_{sqnt}, U_{sqnt})$. For 3C454.3, we
estimated the squint contribution to be small compared to the bumps. 
That the semi-random component of squint might be larger than the mean
squint isn't reasonable.

	\item Another is the squash-like components. The larger of the
two squash-like matrix elements in each of equations \ref{trigeqn1138}
and \ref{trigeqn1} are $\sim 5 \times 10^{-5}$ K$^{-1}$. The unreal
$\sim 0.003$ bumps in $\Delta \tau_{1}(\nu)$ for 3C138 and the unreal $5
\times 10^{-4}$ bumps for 3C454.3 would need bumps in
$Q_{sqsh,cos,true}(\nu)$ of $\sim {0.003 \over 5 \times 10^{-5} \ {\rm
K}^{-1}}=60$ K and $\sim {5 \times 10^{-4}  \over 5 \times 10^{-5} \
{\rm K}^{-1}}=10$ K, respectively. These are  very much larger than the
squint-like contributions---25 times larger for the case 3C454.3.  The
possibility that squash-like contributions are this much larger than
squint ones is unreasonable. 

	\item Finally we have the far-out sidelobes, which are
unmeasurable and unpredictable. 

\end{enumerate}

	By eliminating the other possibilities, we conclude that for
3C138 the observed fake linear polarization results from the polarized
far-out sidelobes. This is the same conclusion we will reach in \S
\ref{comparisonsection} when discussing the empirical squint-like
contribution to $V_{exp,0}(\nu)$.

\section{AN ALTERNATIVE RECIPE FOR DETERMINING THE LEVEL OF INSTRUMENTAL
EFFECTS IN $\tau_{V,2}(\nu)$}

\label{usingqu}

	Our above discussion in \S \ref{empiricalqu} shows two things:
\begin{enumerate}

	\item At Arecibo, polarized sidelobes outside the main beam and
first sidelobe, together with angular structure in the sky, contribute
importantly to the contribution to instrumental polarization;

	\item The effect of polarized sidelobes is expected to be $\sim
10$ times worse in linear than in circular polarization, i.e.\ in Stokes
$(Q,U)$ than in Stokes $V$.

\end{enumerate}

\noindent This leads to use the following alternative recipe for
determining the level of instrumental effects in $\tau_{V,2}(\nu)$.

	First, least-squares fit the Stokes $V$ spectra for squint-like
and squash-like behavior to derive $\tau'_{V,2}(\nu)$. This
eliminates not only the contribution from squint and squash proper
(which, as defined in this paper, come from the main beam and first
sidelobe), but also similar $PA$ behavior arising from the far-out
sidelobes. Similarly, we fit for squint-like behavior in Stokes $(Q,U)$
to derive $\tau'_{Q,1}(\nu),\tau'_{U,1}(\nu)$; we also
gain-correct them to derive $\Delta \tau'_{Q,1}(\nu), \Delta
\tau'_{U,1}(\nu)$. For all three Stokes parameters, the
least-squares fit leaves us with the $PA$-independent portions, which
are the ones of interest.

	Next, we expect the linear polarization to be zero, so we assume
that any departures of $\Delta \tau'_{Q,1}(\nu), \Delta
\tau'_{U,1}(\nu)$ from zero are instrumental, the result of
non-squint-like behavior of the far-out sidelobes. This is an upper
limit because there might, in fact, be true linear polarization. At
Arecibo beam effects in linear polarization are about ten times those in
circular polarization, so calculate the linear polarization
$\tau'_{QU,1}(\nu) = [\Delta \tau'_{Q,1}(\nu)^2 + \Delta
\tau'_{U,1}(\nu)^2]^{1/2}$ and  divide it by ten to estimate the
remaining non-squint-like instrumental effects that remain in the
$\tau'_{V,2}(\nu)$ spectrum.

	In our plots of $\tau_{V,2}'$ we include this alternative
estimate of the uncertainty, along with $\delta \tau'_{V, 2}(\nu)$ as
discussed in \S \ref{stokesvsquintobs}.

\section{SQUINT AND SQUASH CONTRIBUTIONS TO STOKES $V$ EMISSION SPECTRA}

\label{emissionone}

	The above sections concentrate on the uncertainties in the
Stokes $V$ {\it opacity} spectrum $\tau'_V(\nu)$ and its cousins
$\tau'_Q(\nu)$ and $\tau'_U(\nu)$. In principle, we can also derive the
circular polarization of the expected {\it emission} profile
$V_{exp}(\nu)$. Here we address the efficacy of doing this, i.e.\ we
estimate squint-like and squash-like instrumental contributions to
$V_{exp}(\nu)$. We find that we cannot derive reliable values of
$V_{exp}(\nu)$, primarily because of our incomplete sampling of $PA$. 

\subsection{ Empirical evaluation: the example of 3C138}	

\begin{figure}[p!]
\begin{center}
\includegraphics[height=6.0in]{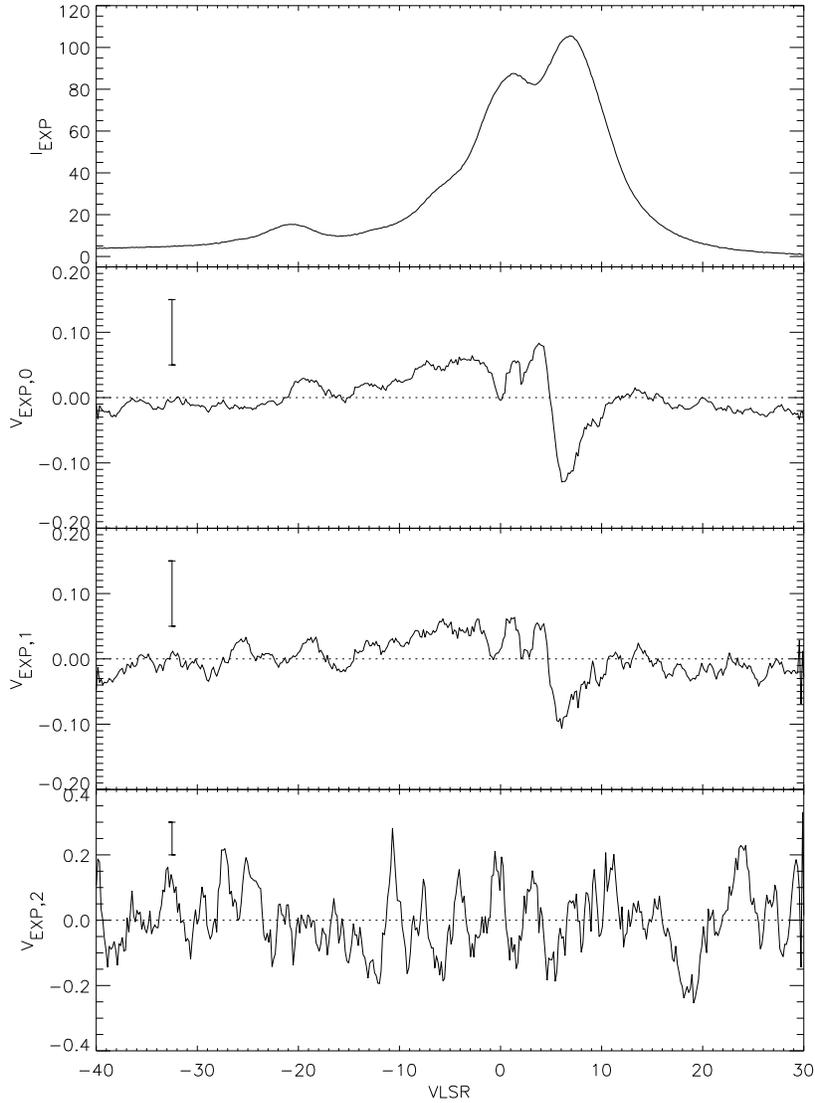}
\end{center}

\caption{This figure is the 3C138 expected emission profile analog of
the opacity spectra in figure \ref{3C138pafig1}. The top panel is the
expected Stokes $I$ emission profile $I_{exp}(\nu)$ (twice the
conventionally-defined brightness temperature). The second panel is
$V_{exp,0}(\nu)$, derived by ignoring squint and squash; the third is 
$V_{exp,1}(\nu)$, derived by including squint but not squash in the fit;
and the fourth is $V_{exp,2}(\nu)$, derived by including both squint and
squash. The bottom three are gain-corrected and boxcar-smoothed by 9
channels for clarity. As shown by the sample errorbar, the vertical
scale on the bottom panel is 2 times larger than on the second and
third panels. \label{3C138pafigexp1.ps} } \end{figure}

\begin{figure}[p!]
\begin{center}
\includegraphics[height=6.75in]{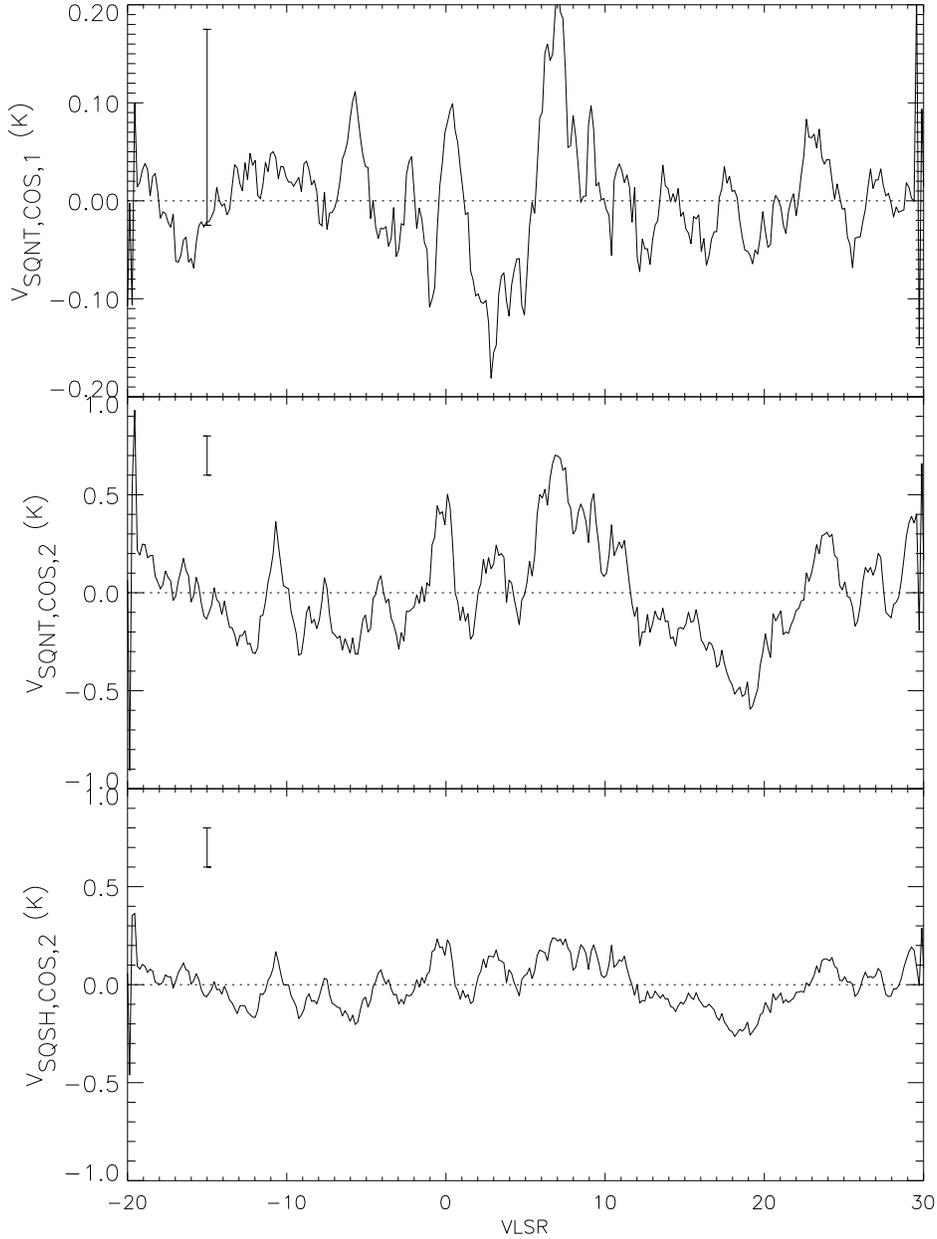}
\end{center}

\caption{The cosine components $V_{sqnt,cos}(\nu)$ and
$V_{sqsh,cos}(\nu)$ for 3C138.  The first (top) panel exhibits $V_{sqnt,
cos, 1}$, derived by solving equation \ref{vfit3} including only the
squint-like terms in equation \ref{trigeq}.  The second and third
exhibit $V_{sqnt, cos,2}$ and $V_{sqsh, cos,2}$ derived by including
both squint-like and squash-like terms. As shown by the sample errorbar,
the vertical scales of the two lower panels are five times that of the
top panel. All spectra are boxcar smoothed by 9 channels to reduce the
noise.  \label{3C138pafigcos} } \end{figure}

\begin{figure}[h!]
\begin{center}
\includegraphics[height=6.75in]{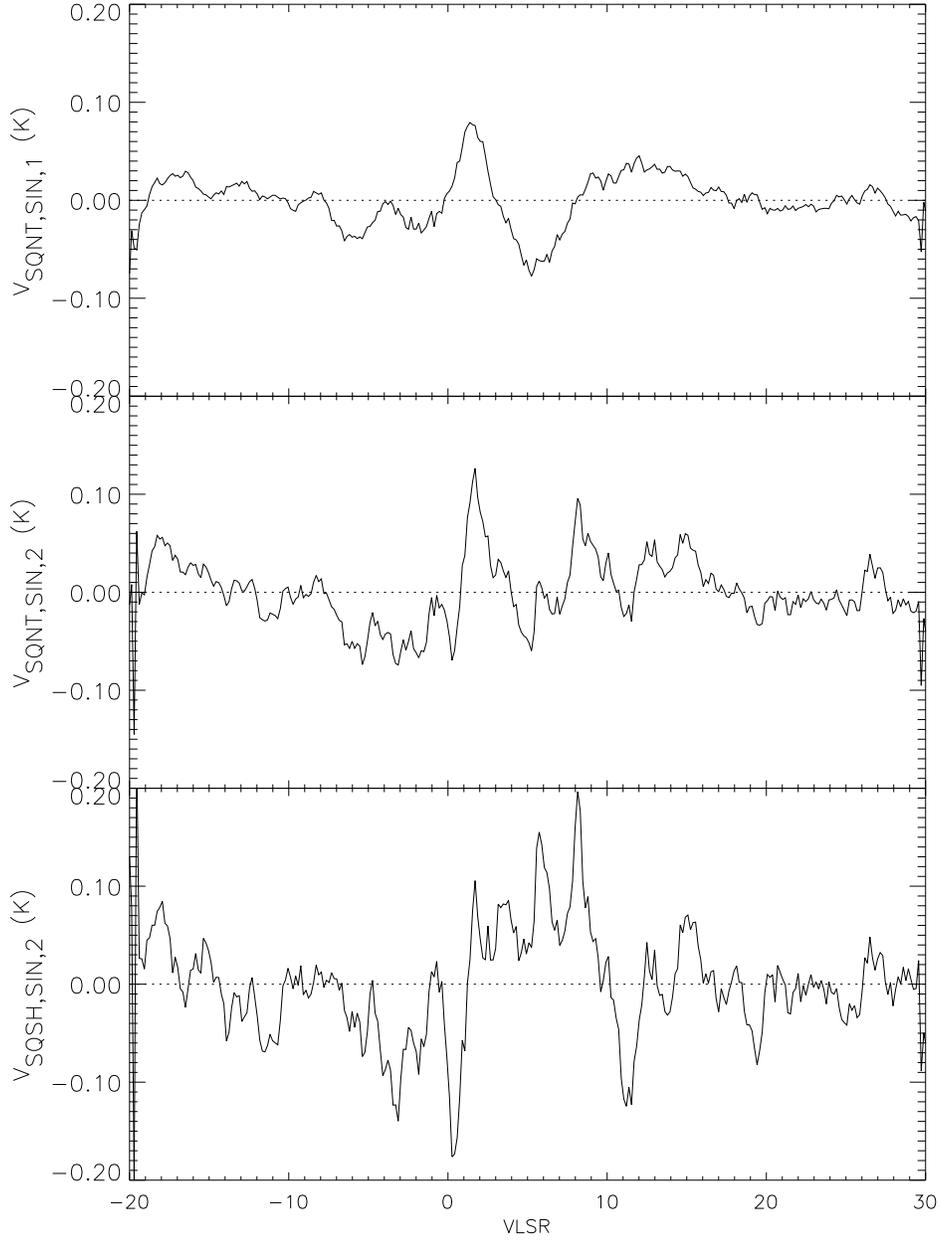}
\end{center}

\caption{The sine components $V_{sqnt,sin}(\nu)$ and $V_{sqsh,sin}(\nu)$
for 3C138.  The first (top) panel exhibits $V_{sqnt, sin,1}$, derived by
solving equation \ref{vfit3} including only the squint-like terms in
equation \ref{trigeq}.  The second and third exhibit $V_{sqnt, sin,2}$
and $V_{sqsh, sin,2}$ derived by including both squint-like and
squash-like terms.   All spectra are boxcar smoothed by 9 channels to
reduce the noise.  \label{3C138pafigsin} } \end{figure}

	As with the polarized opacity spectra, we can derive the
polarized emission spectra for the three cases discussed in \S
\ref{problemevaluationv}, namely ignoring squint and squash, removing
squint only, and removing both squint and squash.  Figure
\ref{3C138pafigexp1.ps} exhibits these three versions $V_{exp,0}(\nu)$,
$V_{exp,1}(\nu)$, and $V_{exp,2}(\nu)$ for 3C138. There are large
differences between the three versions. The spectra change shape and
become noisier as we work our way from $V_{exp,0}(\nu)$ to
$V_{exp,2}(\nu)$. This occurs because of the covariance produced by
incomplete $PA$ coverage, as discussed in \S \ref{problemevaluation}. We
cannot exclude the possibility that most of the contribution to
$V_{exp,0}(\nu)$ is from the squint- and squash-like behaviors, i.e.\
the components of $V_{sqnt,true}(\nu)$ and $V_{sqsh,true}(\nu)$. 

	Figures \ref{3C138pafigcos} and \ref{3C138pafigsin} show the
various measured (not the true) squint- and squash-like contributions.
These contribute to the various versions of $V_{exp}(\nu)$ according to
the relevant matrix elements in the second rows of equations
\ref{trigeqn0138}, \ref{trigeqn1138}, and \ref{trigeqn2138}. As an
example, for $V_{exp,0}(\nu)$, which is derived including neither the
squint-like nor squash-like $PA$ dependence, the ($V_{sqnt, cos,  1},
V_{sqnt, sin,  1}$) components contribute by their values multiplied by
the corresponding matrix elements in equation \ref{trigeqn0138} ($-0.29$
and $-0.21$, respectively). These amount to a fake contributions to
$V_{exp,0}(\nu)$ of $\sim 0.04$ K. This is comparable to the difference
$(V_{exp,0}(\nu) - V_{exp,1}(\nu))$.

	For $V_{exp,1}(\nu)$, derived including only the squint-like
$PA$ dependence, the matrix elements drop to $< 0.005$ (denoted by ``0''
in equation \ref{trigeqn1138}), making the squint-like contribution
negligible. One is tempted to think that the remaining 0.1 K level bumps
in $V_{exp,1}(\nu)$, shown in figure \ref{3C138pafigexp1.ps}, are real.
However, the squash-like matrix element for $V_{sqsh, cos, true}$ is
huge, $-1.33$, so it is conceivable that the 0.1 K bumps are produced by
squash-like behavior; alternatively, they might be produced by far-out
sidelobes whose contribution is neither squint-like nor squash-like. 

	To elucidate these matters we compare the true squint/squash
contributions, which come from only the main beam and first sidelobe,
with the squint-like and squash-like contributions, which come from all
parts of the telescope beam. First we evaluate the true contributions.

\subsection{ Prediction of True Squint and Squash Contributions to
Stokes $V$ Spectra Using Angular Derivatives} \label{twopointfive}

	Here we discuss specifically only {\it true} squint and
squash---namely, those portions of squint-like and squash-like behaviors
that are produced by the primary beam and the first sidelobe. Heiles et
al (2001a) have evaluated the contribution to the Stokes $V$ emission
spectrum from true beam squint and true beam squash interacting with the
first and second spatial derivatives of the spatially extended HI
distribution. We use their formulation to predict this instrumental
(``fake'') contribution, which we denote with the subscript {\it fake}.
In their \S6 Heiles et al (2001a) consult their Figure 14 to find that

\begin{equation} \label{vfake1} 
|V_{fake}| \lesssim \left| 0.015 {dI
\over d\theta} \right| +  \left| 0.025 {d^2I \over d\theta^2} \right| 
\end{equation}

\noindent $I$ and $V$ are antenna temperatures in Kelvins; $\theta$ is
the angle in the sky, units are arcmin. The first term is squint, the
second squash. The $\lesssim$ sign appears because (1) the equation is
approximate; (2) it is an upper limit because it assumes that the
absolute values of the contributions from the first and second spatial
derivatives add arithmetically, while in fact they can cancel; and (3) 
the contributions of each term are periodic in $PA$ or $2PA$, so when
observations are averaged over hour angle the instrumental contributions
partially cancel. 

	Equation~\ref{vfake1} provides the fake $V$ emission spectrum
for a particular position. We could pursue this for the opacity spectra,
also. However, they are derived by ON--OFF observations---in fact, 16 of
them---and the resulting instrumental contributions contain terms in
${d^3I \over d\theta^3}\Delta  \theta^2$ for squint and  ${d^4I \over
d\theta^4}\Delta  \theta^3$ for squash; here $\Delta \theta$ is the
distance between ON and OFF measurements. We cannot evaluate these terms
observationally, so it is not worth discussing opacity spectra.

\subsection{ Comparison of Empirical and Predicted $V$ Opacity Spectra for
the Example of 3C138}

\label{comparisonsection}

\begin{figure}[h!]
\begin{center}
\includegraphics[height=4.0in]{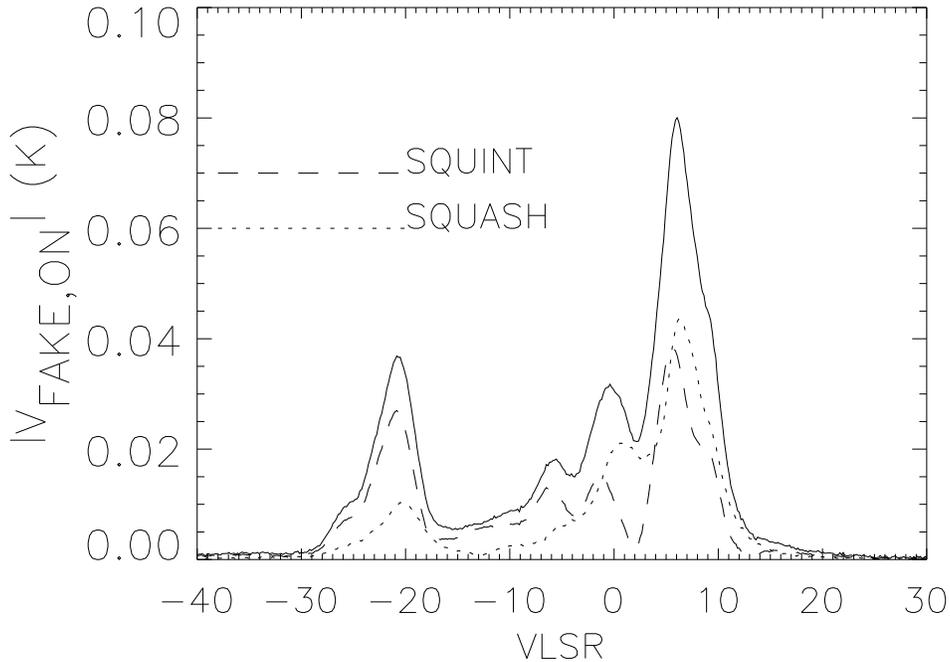}
\end{center}
\caption{The instrumental (``fake'') on-source  emission V profile
$|V_{fake,on}(\nu)|$, from equation \ref{vfake1}. This is the
contribution from true squint and squash, i.e.\ from the primary beam
and first sidelobe only. \label{3C138pafigder_00} } \end{figure}

%

	We denote this predicted spectrum by $V_{exp,fake}(\nu)$ and
show it in Figure \ref{3C138pafigder_00}. There are two distinct parts,
the squint (dashed) and squash (dash-dot), from the first and second
terms in equation \ref{vfake1}, respectively; both are important.
$V_{exp,fake}(\nu)$ has a peak level $\sim 0.08$ K. For comparison, the
top panels of Figures \ref{3C138pafigsin} and \ref{3C138pafigcos} show
the {\it empirically-determined} $V_{sqnt, sin, 1}(\nu)$ and $V_{sqnt,
cos, 1}(\nu)$. They have peak levels $\sim 0.15$ K. This is about 4
times larger than the peak predicted squint. The ratio for some other
sources is larger; for example, for 3C207 the ratio is about 8.
Likewise, the empirically-determined squash-like contributions in figure
\ref{3C138pafigcos} and \ref{3C138pafigsin} (bottom panels) have peak
values $\sim 0.2$ K, about 4-5 times larger than the peak predicted
squash in figure \ref{3C138pafigder_00}.

	We conclude that the empirically-determined squint- and
squash-like contributions are considerably larger than the true squint
and squash expected from the main beam and first sidelobe. The
empirically-determined  ones include not only the true squint and
squash, but also the contribution from far out sidelobes. We conclude
that the far out sidelobes dominate the contribution to the observed
squint and squash. 


	Even weak far out sidelobes can produce these effects. 
They span large angles in the sky. Over these large angles, the HI
angular structure can change considerably; this exaggerates the
contribution of these sidelobes. As a specific example, Heiles et al
{2001a} find that the first sidelobe's contribution to $V_{sqnt, 1}$ is
almost twice that of the main beam, while its contribution to $I$ is
only $\sim {1 \over 3}$ that of the main beam.

	We eliminate the squint-like contribution to $V_{exp}(\nu)$ by
including it in the least squares fit. However, we cannot eliminate the
squash-like component by fitting because of the noise, which results
from the covariance. To obtain accurate results, these instrumental
contributions must be removed with high reliability. This might be
possible by fitting for the squash, as in $V_{exp,1}(\nu)$, but the
excess noise produced by the covariance is prohibitive.  We conclude
that we cannot derive $V_{exp}(\nu)$ for 3C138. 

	Unfortunately, none of our sources exhibits a believable
$V_{exp,1}(\nu)$ profile. In some cases, such as 3C123, $V_{exp,0}(\nu)$
suggests a Zeeman-splitting signal but we have insufficient $PA$
coverage to determine even the squint-like component. We weren't always
able to obtain good $PA$ coverage because of practical considerations
regarding the telescope schedule and our source list. In other cases,
such as 3C138, the squint- and/or squash-like component is disturbingly
large. Telescopes with fewer far-out sidelobes than Arecibo are
desirable, and perhaps even necessary, to determine reliable Zeeman
splitting of emission profiles. 

\subsection{Regarding magnetic fields}

	A magnetic field of $B_{||}$ $\mu$G produces Zeeman splitting
$\delta \nu_Z = 2.8 B_{||}$ Hz, which produces  $V_Z(\nu) \sim {2.8
B_{||} \over \Delta \nu_{FWHM}} I(\nu)$, where $ \Delta \nu_{FWHM}$ is
the half-power linewidth. For a line of width 2 km s$^{-1}$, we have the
uncertainty in magnetic field $\Delta B_{||} \sim 3000 {V(\nu) \over
I(\nu)}$.  For 3C138 we have the peak $I(\nu) \sim 100$ K and the
uncertainty in $V_{exp}(\nu) \sim 0.2$ K, so the uncertainty in
fractional circular polarization is ${\sim 0.002}$. This gives $\Delta
B_{||} \sim 6$ $\mu$G, which is unacceptably large. We conclude that use
of our data for determining $V_{exp,1}(\nu)$, i.e.\ the circular
polarization of {\it emission} profiles, at the levels required for
determining magnetic fields is unwarranted unless the actual magnetic
fields are very high or instrumental effects happen to be unusually
small.

\section{RESULTS}

\label{results}

\subsection{Graphical Results}

	Our Paper I/II survey covered 79 continuum sources. Of these, 61
had detectable CNM. Of these, we had enough integration time for
Zeeman-splitting measurements on 39. Here we add two additional sources,
Tau A (observed at Arecibo) and Cas A (observed years ago at HCRO).
Therefore, we have a total of 41 sources. 

\begin{figure}[p!]
\begin{center}
\includegraphics[height=6.75in]{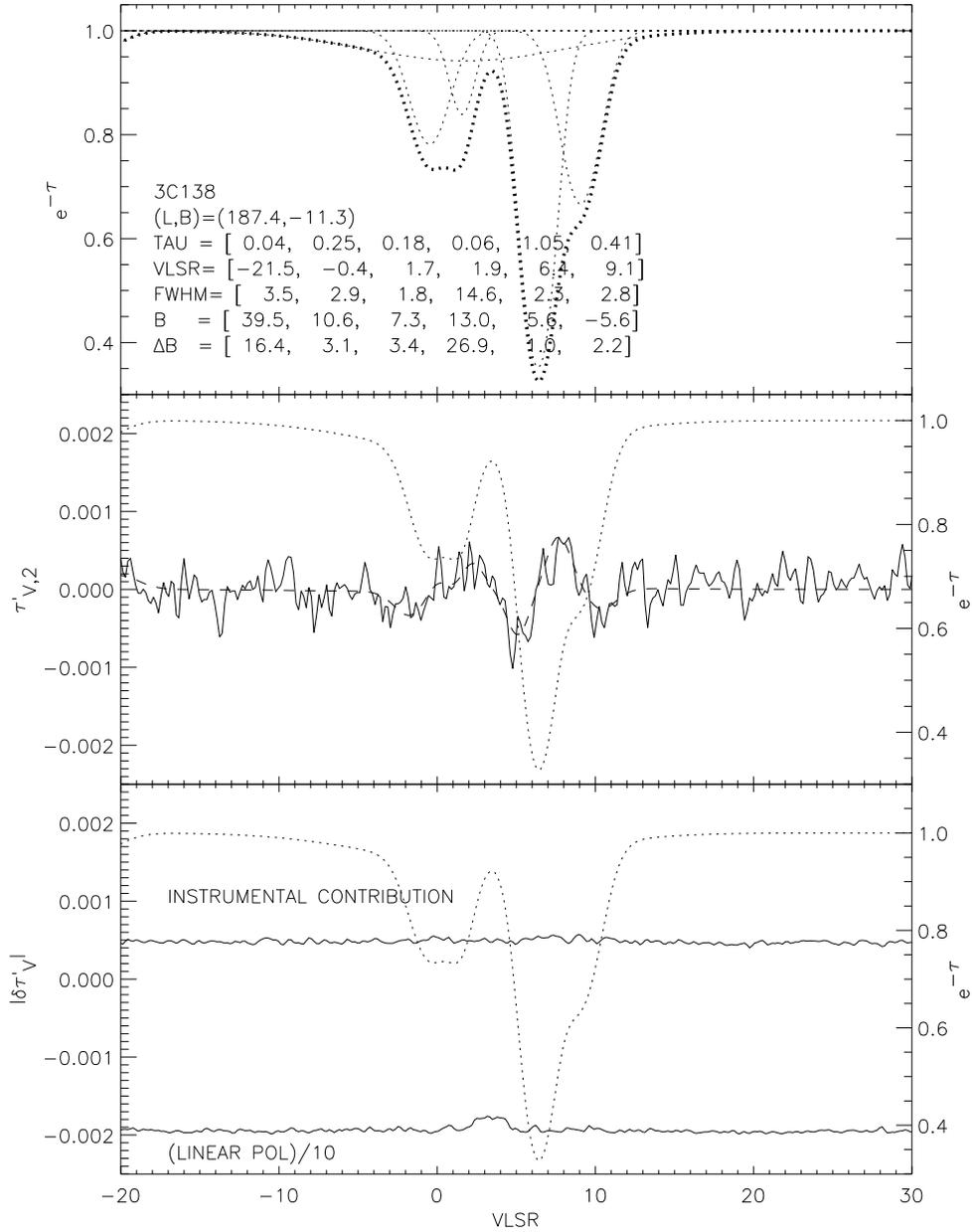}
\end{center} 

\caption{Derived spectra for 3C138; plots for all sources are available
in the electronic edition of {\it The Astrophysical Journal}.  The top
panel exhibits $\tau_0(\nu)$ as the large black dotted line, together
with the fitted Gaussians from Paper I as the light dotted lines. This
panel is annotated with information about the Gaussians. The middle
panel shows $\tau_2'(\nu)$ as the solid line and the least-squares fit as
the dashed line. The bottom panel provides information on possible
instrumental contributions as described in the text.
\label{3C138zmndec02} } \end{figure}

	Figure \ref{3C138zmndec02} presents the data for 3C138 in three
panels. The top panel exhibits $\tau_0(\nu)$ as the large black dotted
line, together with the fitted Gaussians from Paper I as the light
dotted lines. This panel is annotated with information about the
Gaussians. The middle panel shows $\tau_2'(\nu)$ as the solid line, and
the least-squares fit to $B_{||}$ as the dashed line; the fitted field
for each Gaussian is different, and the values are shown in the
annotations of the top panel. The bottom panel provides information on
possible instrumental contributions: the top solid line is the expected
instrumental contribution $\delta \tau_2'(\nu)$ from equation
\ref{trigeqndiff138}, and the bottom solid line shows one-tenth the
linearly polarized profile ${\tau'_{QU,1}(\nu) \over 10}$ as described
in \S \ref{usingqu}. We provide plots equivalent to Figure
\ref{3C138zmndec02} for all sources in the electronic edition of {\it
The Astrophysical Journal}. 

\subsection{Tabular Results for Gaussian Components}

	Typically each source has several CNM components, which we
represent by Gaussians as described in Paper I. Our 41 sources have a
total of 151 components. However, some of these have such large errors
in the derived $B_{||}$ that they are not worth considering. Table
\ref{mkzmntable_mod} lists all Gaussian components for which $\Delta B <
100$ $\mu$G; these number 136. 

	If we define a ``detection'' as $|B_{||}| > 2.5\Delta B_{||}$,
where $\Delta B_{||}$ is the $1\sigma$ uncertainty, and in addition if
we require $\Delta B_{||} < 10$ $\mu$G, then we have 69 components and
22 detections; including results with $\Delta B_{||} < 10$ $\mu$G brings
the 22 up to 26, but we discount these (see below).  If we were to
restrict our discussion to the 22 sure detections, there is little we
could say except to comment on individual sources. However, if we
include the ensemble of results and discuss them statistically, we can
do much more.

	Our results on $B_{||}$ are unique among HI Zeeman-splitting
work in that the errors $\Delta B_{||}$ should be reasonably free of
instrumental contributions, as shown by our discussion in the above
sections. This low instrumental contribution occurs because HI
absorption lines are the results of ON-OFF measurements, which switches
out most of the instrumental contribution. In contrast, other large
Zeeman-splitting surveys (e.g.\ Heiles 1989) examine the HI line in
emission; these are ON measurements, and the instrumental contribution
is a nontrivial portion of the measured results (Heiles 1996). With our
current absorption line measurements, there is no reason not to expect
$\Delta B_{||}$ to be Gaussian distributed. This allows us to use
standard statistical techniques to explore the distribution of field
strength and its correlation with other physical parameters beyond the
small-number statistics of secure detections. 

\subsection{A Good Statistical Sample of Tabular Results}

	In a later paper we will perform the detailed statistical
analyses on our results. In preparation for this work, here we apply
additional criteria to discard an additional selected group of Gaussian
components. We discard these for two reasons. One is to avoid
cluttering plots. The other is because for some components the errors
$\Delta B_{||}$ are likely to depart from the Gaussian distribution;
this can happen  in profiles affected by component blending, especially
when they are noisy.

	We discard Zeeman-splitting detections that satisfy any one of
the following criteria: \begin{enumerate}
 
	\item The uncertainty $\Delta B_{||} > 10$ $\mu$G, unless
${|B_{||}| \over \Delta B_{||}} > 2.5$. We regard such points as
outliers because it is highly unlikely for such strong fields to
exist in HI clouds. 

	\item The interpretation of the profile is complicated. This
occurs for some low-latitude profiles. The discarded data include 3C154,
3C167, 4C22.12, and T0629+10. 

	\item The result is suspicious. This is a subjective judgment
based on the combination of signal/noise, profile complexity, and the
presence of other blended strong components in the same profile that
might have reliable detections. These cases, in which our subjective
judgment creates criteria, include the following:  3C192; 3C274.1; the
18 and 25 km s$^{-1}$ components of 3C410; the --7, 4, and 11 km
s$^{-1}$ components of 3C78; and P0531+19. 

\end{enumerate}

	This leaves us with a total of 69 statistically usable
components. In Table \ref{mkzmntable_mod}, the magnetic fields of the
usable components are in {\bf boldface}.

\subsection{Comparison with previous literature}

	To our knowledge, there exist two sources in our list that have
previous published detections of HI Zeeman splitting. These are the
original discovery of HI Zeeman splitting by Verschuur (1969), who
observed both Tau A and Cas A; and the interferometric study of Schwarz
et al (1986), who studied Cas A. The signs of the Cas A Stokes $V$
spectra  disagree in these two references. From observations of a
calibration helix as well as the 1665 MHz OH maser source W49, we have
determined that Schwarz et al (1986) are correct. Verschuur stated that
his Stokes $V$ was IEEE LHC--RHC, but this appears to be nothing more
than a typographical error. It is not a fundamental sign error because
the signs of his derived $B_{||}$ are correct. Accounting for this
typographical error, the previous two references and the current work
all agree for Cas A, and we agree with Verschuur for Tau A.

\subsection{Yet another source of uncertainty in $B_{||}$}
\label{yetanother}

	In Paper I \S 3, we considered the effect of ordering the
absorbing clouds along the line of sight. This affects the derived spin
temperatures. It also can affect the derived magnetic fields, because
the Stokes $V$ spectrum from a background cloud is weakened by a
foreground one if the velocity profiles overlap.

\begin{figure}[p!]
\begin{center}
\includegraphics[height=4.0in]{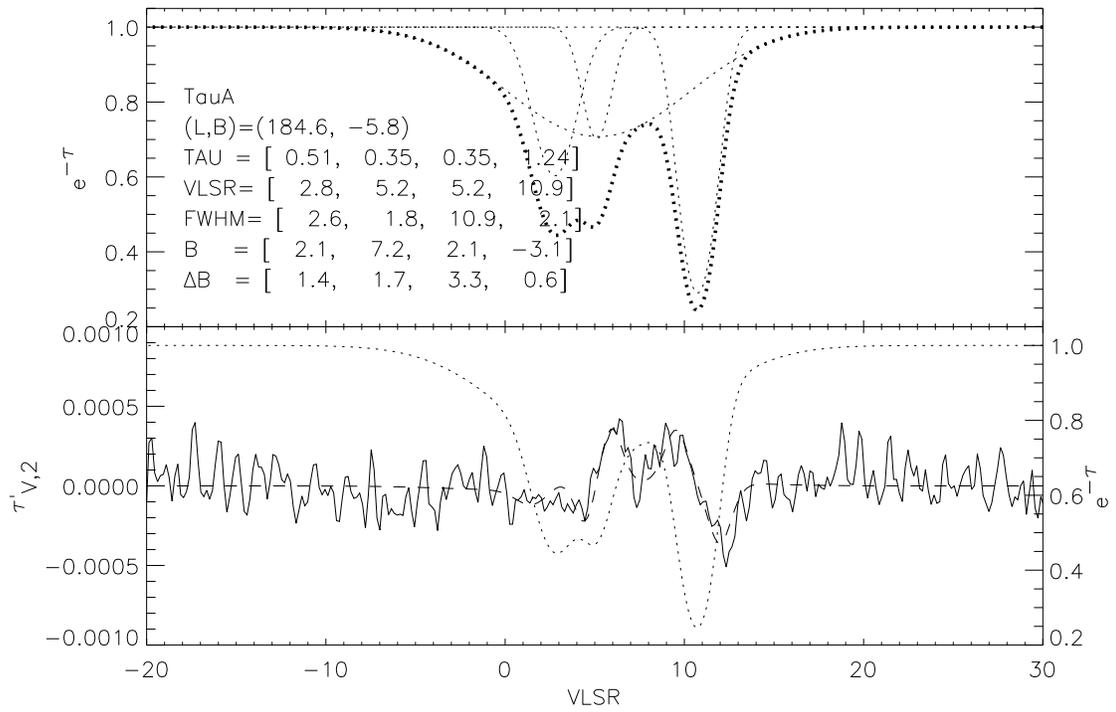}
\end{center} 

\caption{Derived spectra for Tau A; the layout of this figure is
identical to that of Figure \ref{3C138zmndec02} except that here we omit
the bottom panel because Tau A is so strong.  \label {TauAzmndec02}}
\end{figure}

\begin{figure}[p!]
\begin{center}
\includegraphics[height=6.75in]{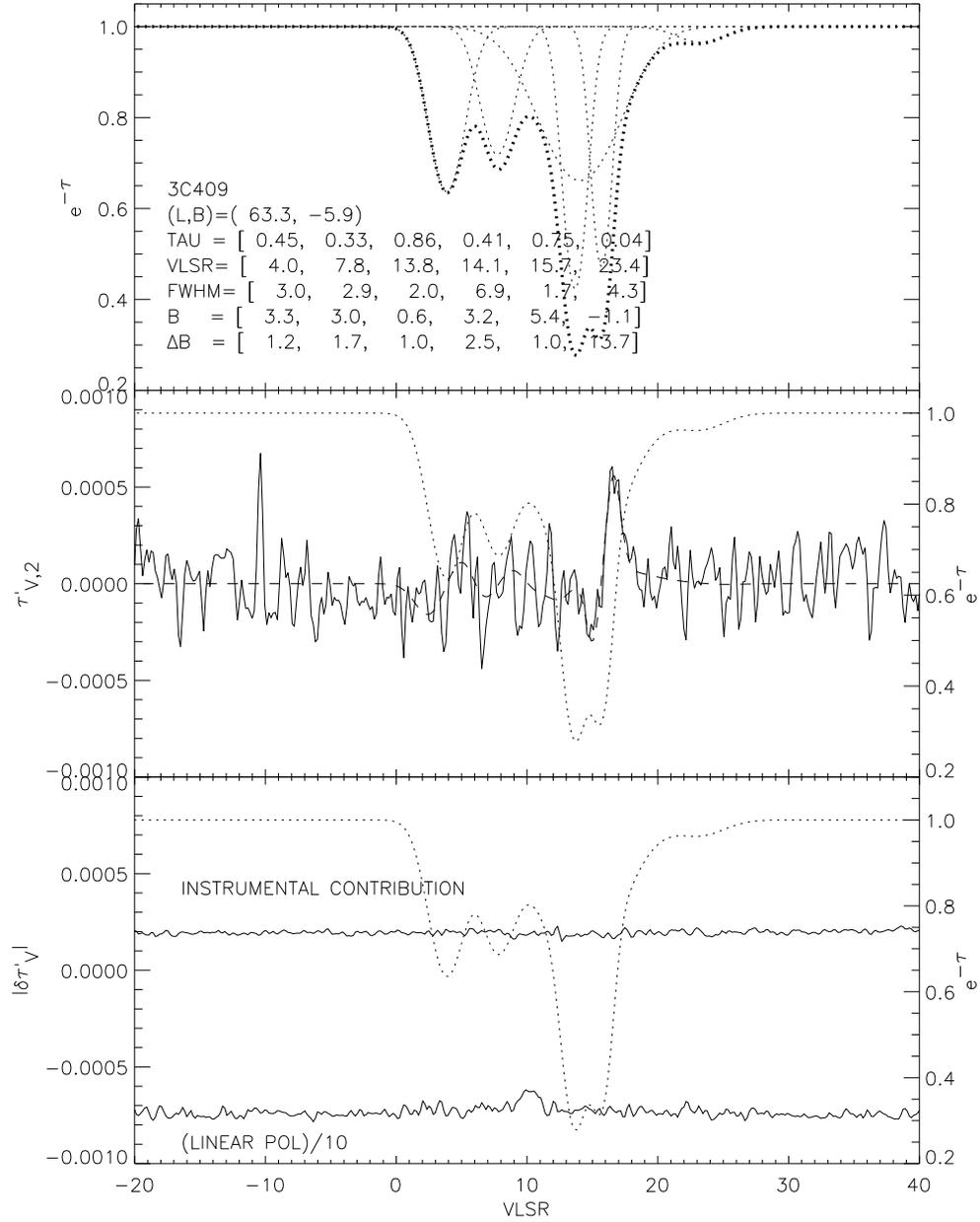}
\end{center} 

\caption{Derived spectra for 3C409; the layout of this figure is
identical to that of Figure \ref{3C138zmndec02}.  \label
{3C409zmndec02}} \end{figure}

\begin{figure}[p!]
\begin{center}
\includegraphics[height=4.0in]{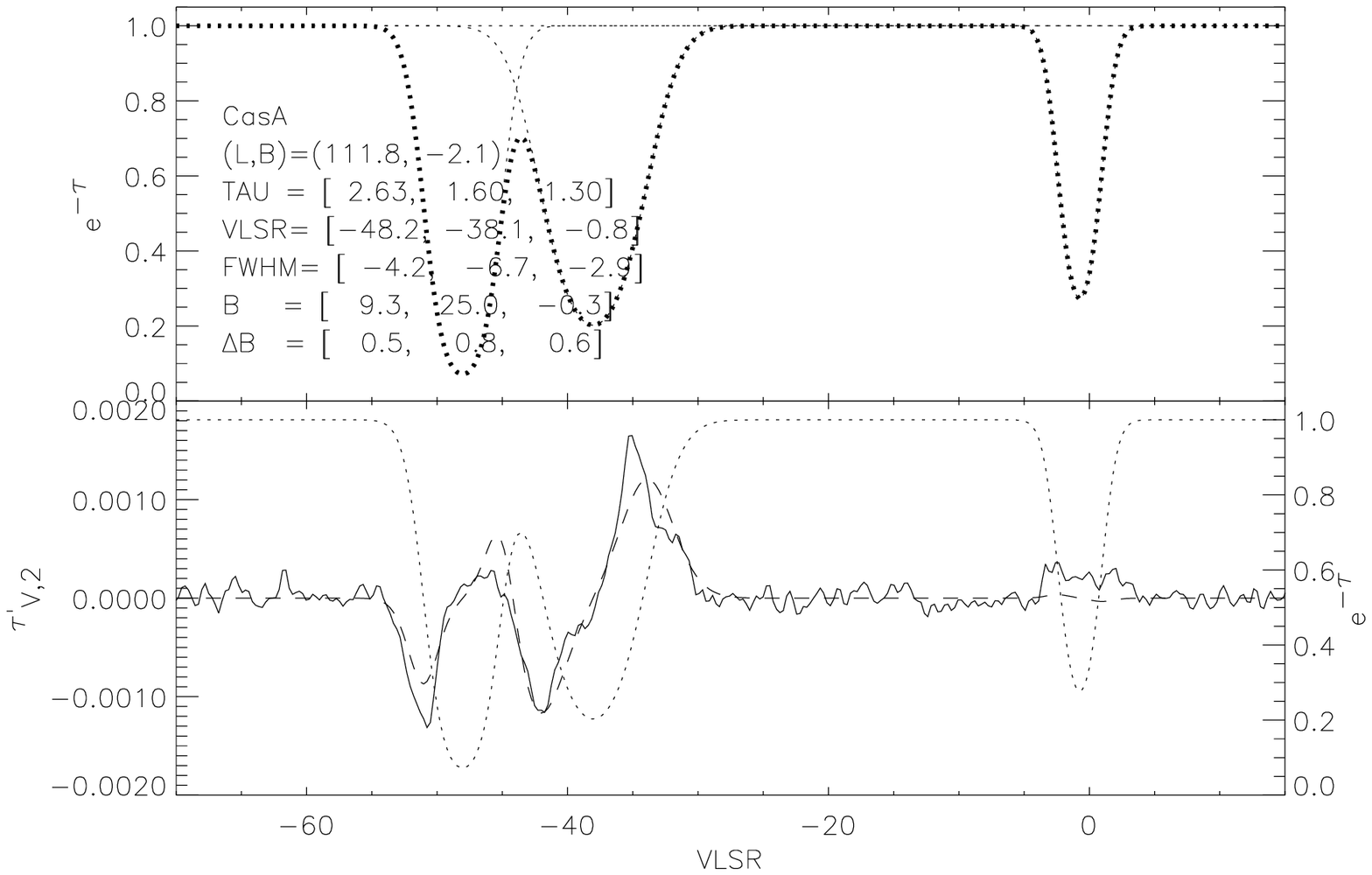}
\end{center} 

\caption{Derived spectra for Cas A; the layout of this figure is
identical to that of Figure \ref{3C138zmndec02} except that here we omit
the bottom panel because Cas A is so strong.  \label {CasAzmndec02}}
\end{figure}

	This affects the derived $B_{||}$ only if there is velocity
overlap and if the opacities are high. For all of our sources, the fits
for the different orderings are visually identical and their variances
differ by insignificant amounts. Thus, as in Paper I, we cannot
determine the line-of-sight ordering. Nevertheless, the ordering affects
the derived magnetic field strengths, as it also does with the spin
temperatures in Paper I. For most sources the differences are smaller
than the $1 \sigma$ uncertainty $\Delta B_{||}$. For three sources,
having a total of four Gaussian components, the differences are larger.
All three sources are at low Galactic latitude where blending is a
problem. We show the spectra for Tau A, 3C409, and Cas A in Figures
\ref{TauAzmndec02}, \ref{3C409zmndec02}, and \ref{CasAzmndec02},
respectively (The Cas A data are from the Hat Creek 85-foot telescope
and are previously unpublished). 

	Table \ref{ordering} lists the four components for which
$B_{||}$ is affected by more than the $1 \sigma$ error $\Delta B_{||}$.
We list three values of field. $B_{||,0}$ is the value from Table
\ref{mkzmntable_mod}; $B_{||,max}$ and $B_{||,min}$ are the minimum and
maximum values obtained from permuting the line-of-sight orderings. 
These differences exceed $\Delta B_{||}$ but are nevertheless modest
fractions of the derived  $B_{||}$. Consequently, we ignore this extra
source of uncertainty both in Table \ref{mkzmntable_mod} and in our
future analyses. 

\section{SUMMARY}

        We discuss the measurement of polarized Stokes parameter 
profiles $(Q,U,V)$ from emission/absorption line observations  toward
background continuum sources. For each of these Stokes  parameters, we
derive the {\it opacity profile} and the {\it emission  profile}
expected if the continuum source were absent.  We give  special emphasis
to the evaluation of instrumental effects.  A  principal motivation for
our study is detection of the Zeeman  effect in Stokes $V$ profiles.
Arecibo suffers with respect to  other telescopes because its large
central blockage produces  large sidelobes. The sidelobes are polarized,
and their  interaction with spatially extended 21-cm emission produces
most  of the instrumental contributions to the polarized profiles. 
These contributions are particularly serious for emission  profiles. But
Arecibo's large collecting area compensates for  this problem when
measuring opacity profiles. As a result, our  Stokes $V$ opacity
profiles are generally reliable so we can make  21-cm Zeeman effect
measurements in the Galactic CNM.

        In \S \ref{stokestheory} we outline the basic theoretical 
concepts involving Stokes parameters in emission/absorption line 
observations. Here we derive the fundamental equation \ref{eqns0}. This 
equation relates the $observed$ Stokes parameter profiles (on and 
off-source) to the physically significant polarized opacity  profiles
and polarized emission profiles. In \S \ref{stokestheory} we also 
explain the contribution of the Zeeman effect to the Stokes $V$ 
profiles.  This contribution is always proportional to the  derivative
of the Stokes $I$ profile (emission or absorption),  regardless of the
line opacity.  Therefore, we can always derive  line-of-sight magnetic
fields by fitting the Stokes $V$ profile  to the derivative of the
Stokes $I$ profile.

        In \S \ref{stokespractice} we present equation \ref{eqnten}.
This  is the fundamental equation that we least squares fit 
(independently for each spectral channel) to derive the Stokes  $V$
opacity profile and the Stokes $V$ emission profile for each  source.
This equation includes an instrumental error term which  is extensively
discussed in later sections. In \S  \ref{stokespractice} we also present
the related discussion for Stokes $Q,U$ profiles, including equation
\ref{skymatrix6}. This equation, the analog of equation \ref{eqnten} for
Stokes $V$, includes a rotation matrix to account for the parallactic
angles of the observations.

        Much of the paper is concerned with instrumental effects 
arising from polarized beam structure.  This structure interacts  with
spatially extended line emission.  As a result, instrumental  effects
appear in the polarized emission profiles and, to a much smaller extent,
in the polarized opacity profiles. In \S  \ref{generaldiscussion} we
describe the nature and the effects of  various types of polarized beam
structure.  One type is ``beam  squint''. Beam squint interacts with the
first spatial derivative  of the Stokes $I$ profile to make instrumental
contributions to  the Stokes $Q,U,V$ profiles. Unfortunately, the beam
squint  contribution to the Stokes $V$ profiles can mimic the Zeeman 
effect. If beam squint remains fixed relative to the telescope  feed
system, then its contributions to the polarized Stokes  profiles vary
with the parallactic angle $PA$.  Another type of  polarized beam
structure is``beam squash''. Beam squash interacts  with the $second$
derivatives of the Stokes $I$ profile to make  instrumental
contributions to Stokes $Q,U,V$. If beam squash  remains fixed relative
to the telescope feed system, then its  contributions vary as $2PA$. For
prime focus telescopes, beam squint is theoretically  expected only for
Stokes $V$, and beam squash is expected only  for Stokes $Q,U$. 
However, Arecibo breaks these rules, having  beam squint and squash in
all polarized Stokes parameters. The  final type of polarized beam
structure arises outside the primary  beam and first sidelobe, in the
``far-out sidelobes''.  Far out  sidelobes at Arecibo are particularly
strong because of the large  aperture blockage.  We define all
contributions to the polarized  Stokes profiles as ``squint-like'' and
``squash-like'' if they are  functions of $PA$ and $2PA$, respectively. 
In practice, squint-like contributions may arise from true beam squint
(in the  primary beam and nearest sidelobe) and, also, from the far-out 
sidelobes.  The same is true for beam squash.

        In \S \ref{problemevaluation} we treat instrumental  squint-like
and squash-like contributions to Stokes $V$ opacity  profiles.  These
profiles are of particular interest for the  Zeeman effect.  Equation
\ref{trigeq} expresses these instrumental  contributions to the
fundamental fitting equation \ref{eqnten} as sinusoidal functions of
$PA$ (squint-like) and $2PA$ (squash-like). We fit equation \ref{eqnten}
in three ways: (a) with neither  squint-like nor squash-like
contributions included, (b) with  squint-like contributions only and (c)
with both squint and  squash-like contributions. Results of these three
types of fits  are Stokes $V$ opacity profiles and emission profiles
with (a) no  instrumental effects removed, (b) squint-like effects
removed and  (c) both squint- and squash-like effects removed,
respectively.  Fits of type (c) are usually best for the Stokes $V$
opacity  profiles, allowing us to remove squint and squash-like 
contributions accurately. In \S \ref{3C138examplev}, we illustrate these
concepts for 3C138 which has a clearly-detected  Zeeman effect. In this
section, we also develop a matrix  representation of the coupling
coefficients between the fitted  Stokes $V$ opacity profile for a given
source and the squint and  squash-like instrumental contributions.  From
this matrix, we  derive a profile of $\delta \tau'_{V,2}(\nu)$ for each
source.   This profile represents the maximum possible instrumental
contributions to the Stokes $V$ opacity profile after squint and 
squash-like contributions have been removed by the fitting  process.  An
example for 3C138 is shown in the bottom panel of  Figure 9.  Here the
profile of $\delta \tau'_{V, 2}(\nu)$ is insignificant compared to the
Zeeman effect in the Stokes $V$ opacity profile.  For most sources, the
profile $\delta \tau'_{V,2}(\nu)$ is very small compared to that of the
Stokes $V$ opacity  profile, another reason to expect that the latter
are reliably  determined.

        \S \ref{empiricalqu} treats squint-like and squash-like 
contributions to the linear polarization Stokes $Q,U$ opacity  profiles.
Least squares fits for squash-like contributions are  not possible:
squash-like effects have the same $2PA$ dependence  as true linear
polarization, so the two cannot be distinguished.  However, true linear
polarization of the HI opacity profile  should be very small,
particularly for small sources like 3C454.3  (angular size 14
milliarcsec). Therefore, the apparent linear polarization we measure in
the 3C454.3 opacity profile must be  instrumental.  We judge that the
same conclusion holds for all  sources, leading to a second method of
estimating instrumental  effects in Stokes $V$ opacity profiles.  The
Arecibo beam squint  and squash are known to be about 10 times greater
in Stokes $Q,U$  than in Stokes $V$. Therefore, we can estimate
instrumental  polarization in Stokes $V$ opacity profiles by dividing
the  apparent linear polarization in the Stokes $Q,U$ profiles by 10.  
Since squint and squash-like effects have already been removed  from
Stokes $V$ opacity profiles, this estimate applies to  instrumental
effects (including those from the far-out sidelobes)  that are not a
function of $PA$ or $2PA$.  This technique is  described in  \S
\ref{usingqu}, and an illustrative profile for  3C138 is shown in the
bottom panel of Figure 9.

        Our final discussion of instrumental effects (\S 
\ref{emissionone}) addresses the reliability of the Stokes $V$  emission
profiles.  For these profiles, fits of type (c) usually  suffer from
large covariance between the $PA$ and $2PA$-dependent  fit parameters. 
As a result, the fit parameters are poorly  determined, and the type (c)
fits yield very noisy Stokes $V$  emission profiles.  An example for
3C138 is shown in the bottom  panel of Figure 5. In effect, we are
unable to remove reliably  the squint- and squash-like contributions
from the Stokes $V$  emission profiles so we cannot derive the Zeeman
effect from  them. Covariance between the fit parameters would be much
less if  we had a larger range of $PA$ included in the data sets for our
 sources.  In \S \ref{emissionone} we also independently evaluate  the
contributions of $true$ beam squint and squash (from the main  beam and
the first sidelobe) to the Stokes $V$ emission profiles.   We compare
these contributions to the empirically-fitted squint  and squash-like
contributions (from the main beam and all  sidelobes). The difference
between these two is the contribution  from the far-out sidelobes alone.
We find that the far-out  sidelobes contribute most of the instrumental
effects at Arecibo.   We expect these contributions to be mainly squint
or squash-like, but there might also other contributions that do not
depend on $PA$ or $2PA$; such contributions contribute additional
uncertainties. {\bf NOTE THIS CHANGE FROM ORIGINAL!}  For all these
reasons, Zeeman effect results derived from Arecibo Stokes $V$ emission
profiles are unreliable.

        \S \ref{results} presents the Stokes $I$ and $V$ opacity 
profiles for all sources, along with profiles of possible  instrumental
effects in the latter.  Profiles for 3C138 are shown  in Figure 9,
profiles for other sources others are provided  electronically. This
section also presents a tabular list of  parameters, including
line-of-sight magnetic field strengths, for  the CNM Gaussian components
in the Stokes $I$ opacity profiles.  We select a sample 69 Gaussian
components for which $\Delta B_{||}$ (uncertainties in the derived
$B_{||}$) should be Gaussian distributed and, additionally, for which
$\Delta B_{||} < 10$ $\mu$G. {\bf NOTE THIS CHANGE FROM ORIGINAL!}  We
will subject this sample to a future statistical analysis of magnetic
field strengths in the CNM.  In \S  \ref{results}, we also discuss
another source of uncertainty in  derived magnetic field strengths, the
unknown sequential order along the  line-of-sight of the various CNM
velocity components.  However,  this source of error has no significant
effect upon the  qualitative or statistical properties of the magnetic
field  measurements.

\acknowledgements

	This work was supported in part by NSF grants AST-9530590,
AST-0097417, and AST-9988341; and by the NAIC.



\begin{deluxetable}{ccccccccc}
\tabletypesize{\scriptsize}
\tablewidth{0pc}
\tablecaption{Table of Gaussian Fit Parameters Having
        $\Delta B_{||} < 100$ $\mu$G \label{mkzmntable_mod} }
\tablehead{
\colhead{ $T_B$ } &
\colhead{ $ \tau $ } &
\colhead{ $ V_{LSR} $ } &
\colhead{ $\Delta V$ } &
\colhead{ $ T_s$ } &
\colhead{ $ T_{kmax}$ } &
\colhead{ $ N(HI)_{20}$ } &
\colhead{ $ B_{||}$ } &
\colhead{ $ (l/b/$SOURCE) }
}
\startdata
   34.34& $   1.666\pm   0.007$ & $     7.6\pm     0.0$ & $    4.48\pm    0.01$ & $      42.34\pm   9.96$ &   438&  6.15&$\mathbf{  -1.6\pm 1.9 }$ &190.4/$-27.4$/3C120 \\
   19.32& $   0.736\pm   0.010$ & $     6.2\pm     0.0$ & $    1.33\pm    0.02$ & $      37.08\pm   6.86$ &    38&  0.71&$\mathbf{   5.2\pm 3.8 }$ &190.4/$-27.4$/3C120 \\
   20.65& $   0.634\pm   0.008$ & $    10.2\pm     0.0$ & $    1.91\pm    0.02$ & $      43.96\pm   3.95$ &    79&  1.04&$\mathbf{   3.6\pm 2.8 }$ &190.4/$-27.4$/3C120 \\
   80.96& $   0.596\pm   0.026$ & $     4.7\pm     0.0$ & $    2.32\pm    0.08$ & $     180.36\pm 115.21$ &   117&  4.86&$\mathbf{   4.8\pm 4.2 }$ &170.6/$-11.7$/3C123 \\
   13.50& $   1.606\pm   0.026$ & $     4.4\pm     0.0$ & $    4.79\pm    0.02$ & $      16.88\pm   7.44$ &   501&  2.53&$\mathbf{  -2.6\pm 1.0 }$ &170.6/$-11.7$/3C123 \\
    2.51& $   0.063\pm   0.002$ & $   -19.6\pm     0.0$ & $    3.14\pm    0.11$ & $      40.93\pm   6.20$ &   215&  0.16&$\mathbf{   1.2\pm 8.6 }$ &170.6/$-11.7$/3C123 \\
    0.18& $   0.008\pm   0.001$ & $   -57.8\pm     0.5$ & $    5.62\pm    1.09$ & $      22.15\pm  40.62$ &   691&  0.02&$ -11.4\pm86.9 $ &170.6/$-11.7$/3C123 \\
    0.57& $   0.032\pm   0.002$ & $   -72.9\pm     0.1$ & $    4.28\pm    0.24$ & $      17.79\pm  10.26$ &   400&  0.05&$  -4.7\pm19.3 $ &170.6/$-11.7$/3C123 \\
    0.00& $   0.007\pm   0.002$ & $    20.0\pm     0.5$ & $    3.88\pm    1.09$ & $       0.00\pm   0.00$ &   328&  0.00&$  55.1\pm86.1 $ &170.6/$-11.7$/3C123 \\
   39.30& $   2.152\pm   0.034$ & $     5.3\pm     0.0$ & $    4.24\pm    0.04$ & $      44.47\pm   8.06$ &   392&  7.90&$\mathbf{   0.7\pm 8.4 }$ &171.4/$ -7.8$/3C131 \\
   21.36& $   0.321\pm   0.006$ & $    -2.3\pm     0.1$ & $    6.95\pm    0.21$ & $      77.79\pm   5.43$ &  1057&  3.38&$   2.8\pm36.2 $ &171.4/$ -7.8$/3C131 \\
   19.68& $   0.260\pm   0.006$ & $    13.0\pm     0.1$ & $    5.68\pm    0.23$ & $      85.86\pm   9.64$ &   705&  2.47&$\mathbf{   9.7\pm 5.5 }$ &178.9/$-12.5$/3C132 \\
   43.75& $   1.542\pm   0.028$ & $     8.1\pm     0.0$ & $    2.42\pm    0.03$ & $      55.66\pm  14.67$ &   128&  4.05&$\mathbf{   4.2\pm 1.0 }$ &178.9/$-12.5$/3C132 \\
    9.04& $   0.351\pm   0.007$ & $     1.8\pm     0.1$ & $    5.43\pm    0.12$ & $      30.52\pm  11.08$ &   643&  1.13&$\mathbf{  -4.0\pm 3.8 }$ &178.9/$-12.5$/3C132 \\
   30.18& $   1.532\pm   0.021$ & $     8.0\pm     0.0$ & $    2.51\pm    0.02$ & $      38.50\pm  13.68$ &   138&  2.89&$\mathbf{   5.8\pm 1.1 }$ &177.7/$ -9.9$/3C133 \\
   25.98& $   0.891\pm   0.021$ & $     3.7\pm     0.0$ & $    2.77\pm    0.06$ & $      44.04\pm  17.06$ &   167&  2.12&$\mathbf{  -0.3\pm 1.7 }$ &177.7/$ -9.9$/3C133 \\
   15.35& $   0.262\pm   0.006$ & $    -0.2\pm     0.2$ & $    6.17\pm    0.31$ & $      66.60\pm  11.51$ &   831&  2.10&$\mathbf{  -9.5\pm 6.3 }$ &177.7/$ -9.9$/3C133 \\
    1.16& $   0.064\pm   0.009$ & $   -27.6\pm     0.2$ & $    2.84\pm    0.46$ & $      18.79\pm   6.96$ &   176&  0.07&$  15.1\pm14.2 $ &177.7/$ -9.9$/3C133 \\
   12.83& $   0.060\pm   0.008$ & $   -29.5\pm     0.4$ & $    8.45\pm    0.63$ & $     219.27\pm   5.45$ &  1559&  2.17&$ -41.8\pm25.0 $ &177.7/$ -9.9$/3C133 \\
   25.99& $   1.046\pm   0.008$ & $     6.4\pm     0.0$ & $    2.30\pm    0.02$ & $      40.07\pm  12.30$ &   115&  1.87&$\mathbf{   5.6\pm 1.0 }$ &187.4/$-11.3$/3C138 \\
   15.70& $   0.406\pm   0.005$ & $     9.1\pm     0.0$ & $    2.81\pm    0.06$ & $      47.02\pm  11.54$ &   172&  1.05&$\mathbf{  -5.6\pm 2.2 }$ &187.4/$-11.3$/3C138 \\
    8.48& $   0.176\pm   0.014$ & $     1.6\pm     0.1$ & $    1.84\pm    0.09$ & $      52.62\pm  11.03$ &    73&  0.33&$\mathbf{   7.3\pm 3.4 }$ &187.4/$-11.3$/3C138 \\
   11.82& $   0.247\pm   0.006$ & $    -0.5\pm     0.1$ & $    2.86\pm    0.12$ & $      54.03\pm   9.60$ &   178&  0.74&$\mathbf{  10.6\pm 3.1 }$ &187.4/$-11.3$/3C138 \\
   21.94& $   0.060\pm   0.004$ & $     1.8\pm     0.2$ & $   14.64\pm    0.48$ & $     379.12\pm  23.44$ &  4683&  6.44&$  13.0\pm26.9 $ &187.4/$-11.3$/3C138 \\
    3.78& $   0.038\pm   0.002$ & $   -21.5\pm     0.1$ & $    3.45\pm    0.21$ & $     101.52\pm   5.08$ &   260&  0.26&$  39.5\pm16.4 $ &187.4/$-11.3$/3C138 \\
   44.11& $   2.362\pm   0.045$ & $     7.0\pm     0.0$ & $    3.13\pm    0.03$ & $      48.69\pm  15.55$ &   214&  7.02&$\mathbf{  -8.3\pm 1.3 }$ &197.6/$-14.5$/3C142.1 \\
    4.22& $   0.203\pm   0.007$ & $    13.4\pm     0.1$ & $    4.12\pm    0.17$ & $      23.03\pm  10.75$ &   370&  0.37&$\mathbf{   7.2\pm 7.4 }$ &197.6/$-14.5$/3C142.1 \\
    9.51& $   0.083\pm   0.007$ & $    22.4\pm     0.1$ & $    3.39\pm    0.32$ & $     119.08\pm  10.47$ &   251&  0.65&$ -23.3\pm14.9 $ &197.6/$-14.5$/3C142.1 \\
    0.90& $   0.101\pm   0.007$ & $    -9.2\pm     0.1$ & $    3.24\pm    0.26$ & $       9.37\pm  13.82$ &   229&  0.06&$   0.8\pm12.1 $ &197.6/$-14.5$/3C142.1 \\
   30.83& $   0.919\pm   0.014$ & $     1.8\pm     0.0$ & $    2.63\pm    0.07$ & $      51.28\pm  18.44$ &   151&  2.42&$  -9.8\pm 1.7 $ &185.6/$  4.0$/3C154 \\
    6.05& $   0.292\pm   0.019$ & $    -2.1\pm     0.0$ & $    1.32\pm    0.10$ & $      23.87\pm  18.23$ &    37&  0.18&$  -6.0\pm 4.0 $ &185.6/$  4.0$/3C154 \\
   24.81& $   0.709\pm   0.014$ & $    -2.9\pm     0.0$ & $    4.45\pm    0.06$ & $      48.85\pm  14.40$ &   433&  3.01&$  -5.1\pm 2.4 $ &185.6/$  4.0$/3C154 \\
   19.70& $   0.479\pm   0.007$ & $     5.0\pm     0.1$ & $    3.48\pm    0.10$ & $      51.78\pm  15.58$ &   263&  1.68&$  -6.3\pm 2.8 $ &185.6/$  4.0$/3C154 \\
   10.83& $   0.413\pm   0.006$ & $    10.6\pm     0.0$ & $    2.12\pm    0.03$ & $      31.98\pm   9.17$ &    98&  0.55&$   0.1\pm 2.0 $ &185.6/$  4.0$/3C154 \\
   11.61& $   0.068\pm   0.003$ & $   -23.7\pm     0.1$ & $    4.35\pm    0.22$ & $     176.60\pm   5.51$ &   412&  1.02&$  -3.9\pm14.3 $ &185.6/$  4.0$/3C154 \\
   38.18& $   0.252\pm   0.008$ & $    22.6\pm     0.4$ & $   21.71\pm    0.98$ & $     171.23\pm   6.90$ & 10301& 18.26&$ -40.5\pm33.1 $ &207.3/$  1.2$/3C167 \\
   50.50& $   0.941\pm   0.024$ & $    42.2\pm     0.1$ & $    8.01\pm    0.23$ & $      82.85\pm  12.73$ &  1403& 12.16&$  39.2\pm 7.7 $ &207.3/$  1.2$/3C167 \\
    8.60& $   0.386\pm   0.032$ & $    49.3\pm     0.1$ & $    2.09\pm    0.19$ & $      26.87\pm   5.61$ &    95&  0.42&$ -24.6\pm 7.6 $ &207.3/$  1.2$/3C167 \\
   16.71& $   0.669\pm   0.011$ & $    -8.9\pm     0.0$ & $    2.43\pm    0.03$ & $      34.25\pm   7.89$ &   129&  1.09&$\mathbf{  -1.2\pm 1.5 }$ &118.6/$-52.7$/3C18 \\
    8.15& $   0.183\pm   0.018$ & $    -5.7\pm     0.1$ & $    3.99\pm    0.25$ & $      48.75\pm   9.41$ &   347&  0.69&$\mathbf{  12.8\pm 7.0 }$ &118.6/$-52.7$/3C18 \\
   19.42& $   0.075\pm   0.019$ & $    -6.7\pm     0.2$ & $    8.68\pm    0.70$ & $     267.80\pm   5.28$ &  1648&  3.41&$  26.5\pm21.6 $ &118.6/$-52.7$/3C18 \\
    5.39& $   0.068\pm   0.002$ & $     8.0\pm     0.1$ & $    4.30\pm    0.12$ & $      82.14\pm   3.76$ &   403&  0.47&$ -28.1\pm12.0 $ &197.9/$ 26.4$/3C192 \\
    5.03& $   0.298\pm   0.004$ & $    15.4\pm     0.0$ & $    2.43\pm    0.03$ & $      19.53\pm   2.66$ &   129&  0.28&$\mathbf{  -1.9\pm 2.2 }$ &213.0/$ 30.1$/3C207 \\
    5.46& $   0.250\pm   0.002$ & $     4.2\pm     0.0$ & $    5.25\pm    0.05$ & $      24.65\pm   4.74$ &   602&  0.63&$\mathbf{  -3.2\pm 3.7 }$ &213.0/$ 30.1$/3C207 \\
    5.78& $   0.313\pm   0.003$ & $     4.0\pm     0.0$ & $    1.32\pm    0.01$ & $      21.53\pm   1.30$ &    37&  0.17&$\mathbf{  -1.2\pm 4.9 }$ &219.9/$ 44.0$/3C225a \\
    9.17& $   0.745\pm   0.002$ & $     3.6\pm     0.0$ & $    1.25\pm    0.00$ & $      17.44\pm   1.80$ &    34&  0.32&$\mathbf{  -1.3\pm 1.1 }$ &220.0/$ 44.0$/3C225b \\
    1.21& $   0.027\pm   0.001$ & $   -28.0\pm     0.1$ & $    4.78\pm    0.13$ & $      45.40\pm   3.28$ &   499&  0.11&$   3.6\pm43.4 $ &220.0/$ 44.0$/3C225b \\
    2.79& $   0.047\pm   0.001$ & $   -37.9\pm     0.1$ & $    2.52\pm    0.11$ & $      61.11\pm   1.97$ &   138&  0.14&$ -15.7\pm20.2 $ &220.0/$ 44.0$/3C225b \\
    0.53& $   0.033\pm   0.001$ & $   -40.6\pm     0.1$ & $    2.07\pm    0.12$ & $      16.64\pm   3.05$ &    93&  0.02&$   1.8\pm26.1 $ &220.0/$ 44.0$/3C225b \\
    4.47& $   0.398\pm   0.001$ & $     1.9\pm     0.0$ & $    1.19\pm    0.00$ & $      13.61\pm   0.26$ &    31&  0.13&$\mathbf{  -0.7\pm 1.1 }$ &232.1/$ 46.6$/3C237 \\
    1.39& $   0.005\pm   0.000$ & $    -3.0\pm     0.1$ & $    2.48\pm    0.16$ & $     255.60\pm  10.79$ &   134&  0.07&$  45.6\pm92.9 $ &232.1/$ 46.6$/3C237 \\
    0.81& $   0.018\pm   0.000$ & $    -6.3\pm     0.0$ & $    2.37\pm    0.04$ & $      44.43\pm   3.19$ &   122&  0.04&$ -29.6\pm24.6 $ &289.9/$ 64.4$/3C273 \\
    3.80& $   0.102\pm   0.001$ & $    -1.6\pm     0.0$ & $    2.94\pm    0.04$ & $      39.15\pm   2.41$ &   189&  0.23&$  44.9\pm13.1 $ &269.9/$ 83.2$/3C274.1 \\
   17.93& $   0.620\pm   0.003$ & $    -3.7\pm     0.0$ & $    1.75\pm    0.01$ & $      38.81\pm   3.37$ &    66&  0.82&$\mathbf{  -2.7\pm 1.3 }$ & 38.5/$ 60.2$/3C310 \\
    2.86& $   0.061\pm   0.001$ & $     0.6\pm     0.1$ & $    5.11\pm    0.13$ & $      48.36\pm   4.76$ &   571&  0.29&$   9.2\pm17.8 $ & 38.5/$ 60.2$/3C310 \\
   24.03& $   0.784\pm   0.011$ & $    -4.2\pm     0.0$ & $    2.15\pm    0.02$ & $      44.23\pm   2.20$ &   100&  1.45&$\mathbf{  -0.1\pm 1.1 }$ & 39.3/$ 58.5$/3C315 \\
    8.30& $   0.146\pm   0.004$ & $     1.6\pm     0.1$ & $    4.41\pm    0.15$ & $      61.00\pm  15.11$ &   425&  0.77&$\mathbf{   3.9\pm 6.3 }$ & 39.3/$ 58.5$/3C315 \\
   13.05& $   0.482\pm   0.013$ & $    -6.0\pm     0.0$ & $    1.77\pm    0.03$ & $      34.12\pm   4.47$ &    68&  0.57&$\mathbf{  -0.2\pm 4.1 }$ & 30.0/$ 54.8$/3C318 \\
   15.55& $   0.300\pm   0.011$ & $    -5.0\pm     0.0$ & $    3.36\pm    0.04$ & $      60.07\pm   5.81$ &   246&  1.18&$\mathbf{  -4.6\pm 7.6 }$ & 30.0/$ 54.8$/3C318 \\
   17.01& $   0.993\pm   0.010$ & $     0.9\pm     0.0$ & $    2.11\pm    0.02$ & $      27.01\pm   7.96$ &    97&  1.10&$\mathbf{   3.0\pm 1.7 }$ & 37.6/$ 42.3$/3C333 \\
    9.35& $   0.025\pm   0.000$ & $    -4.3\pm     0.0$ & $    8.91\pm    0.10$ & $     379.10\pm   3.26$ &  1733&  1.64&$ -30.9\pm42.9 $ &129.4/$-49.3$/3C33 \\
    2.66& $   0.259\pm   0.003$ & $    -2.2\pm     0.0$ & $    1.65\pm    0.03$ & $      11.65\pm   4.80$ &    59&  0.10&$\mathbf{   1.4\pm 1.5 }$ & 23.0/$ 29.2$/3C348 \\
   14.76& $   0.604\pm   0.004$ & $     0.5\pm     0.0$ & $    2.12\pm    0.01$ & $      32.54\pm   5.82$ &    98&  0.81&$\mathbf{   0.0\pm 0.9 }$ & 23.0/$ 29.2$/3C348 \\
    8.53& $   0.078\pm   0.002$ & $     7.2\pm     0.0$ & $    3.73\pm    0.09$ & $     113.21\pm   1.57$ &   304&  0.64&$\mathbf{   0.5\pm 6.4 }$ & 23.0/$ 29.2$/3C348 \\
   25.87& $   1.209\pm   0.007$ & $     0.0\pm     0.0$ & $    2.80\pm    0.01$ & $      36.89\pm  10.12$ &   170&  2.43&$\mathbf{   4.2\pm 2.0 }$ & 21.1/$ 19.9$/3C353 \\
   15.37& $   0.859\pm   0.006$ & $     2.4\pm     0.0$ & $    1.69\pm    0.01$ & $      26.66\pm  13.08$ &    62&  0.76&$\mathbf{   5.1\pm 2.0 }$ & 21.1/$ 19.9$/3C353 \\
   28.62& $   0.195\pm   0.008$ & $     1.4\pm     0.0$ & $    5.84\pm    0.07$ & $     161.35\pm   4.53$ &   746&  3.59&$   0.6\pm11.8 $ & 21.1/$ 19.9$/3C353 \\
    1.57& $   0.040\pm   0.001$ & $    11.9\pm     0.0$ & $    3.01\pm    0.07$ & $      39.89\pm   5.32$ &   198&  0.09&$ -15.7\pm21.5 $ & 21.1/$ 19.9$/3C353 \\
   11.93& $   0.451\pm   0.002$ & $     3.9\pm     0.0$ & $    3.03\pm    0.02$ & $      32.85\pm   7.01$ &   201&  0.88&$\mathbf{   3.3\pm 1.2 }$ & 63.3/$ -5.9$/3C409 \\
   11.86& $   0.331\pm   0.003$ & $     7.7\pm     0.0$ & $    2.91\pm    0.04$ & $      42.08\pm  10.70$ &   184&  0.79&$\mathbf{   3.0\pm 1.7 }$ & 63.3/$ -5.9$/3C409 \\
   20.64& $   0.753\pm   0.009$ & $    15.7\pm     0.0$ & $    1.69\pm    0.02$ & $      38.99\pm  13.10$ &    62&  0.97&$\mathbf{   5.4\pm 1.0 }$ & 63.3/$ -5.9$/3C409 \\
   22.40& $   0.861\pm   0.010$ & $    13.7\pm     0.0$ & $    1.96\pm    0.02$ & $      38.80\pm  18.47$ &    83&  1.28&$\mathbf{   0.6\pm 1.0 }$ & 63.3/$ -5.9$/3C409 \\
   10.44& $   0.413\pm   0.011$ & $    14.0\pm     0.0$ & $    6.90\pm    0.13$ & $      30.85\pm  11.38$ &  1040&  1.71&$\mathbf{   3.2\pm 2.5 }$ & 63.3/$ -5.9$/3C409 \\
    5.17& $   0.035\pm   0.001$ & $    23.3\pm     0.1$ & $    4.27\pm    0.20$ & $     148.28\pm  13.67$ &   399&  0.44&$  -1.1\pm13.7 $ & 63.3/$ -5.9$/3C409 \\
   52.84& $   2.214\pm   0.030$ & $     7.8\pm     0.0$ & $    2.77\pm    0.05$ & $      59.33\pm  24.84$ &   167&  7.08&$\mathbf{   1.4\pm 1.0 }$ & 69.2/$ -3.8$/3C410 \\
   31.74& $   0.688\pm   0.009$ & $    11.3\pm     0.1$ & $    3.55\pm    0.11$ & $      63.79\pm  22.10$ &   276&  3.04&$\mathbf{   0.2\pm 2.1 }$ & 69.2/$ -3.8$/3C410 \\
   13.06& $   0.369\pm   0.061$ & $    -0.7\pm     0.2$ & $    3.51\pm    0.21$ & $      42.30\pm  13.68$ &   269&  1.07&$\mathbf{   4.4\pm 3.0 }$ & 69.2/$ -3.8$/3C410 \\
   26.67& $   0.654\pm   0.025$ & $     2.6\pm     0.2$ & $    4.33\pm    0.34$ & $      55.56\pm  22.19$ &   409&  3.06&$\mathbf{   5.2\pm 2.3 }$ & 69.2/$ -3.8$/3C410 \\
    0.00& $   0.118\pm   0.004$ & $    17.8\pm     0.1$ & $    4.81\pm    0.25$ & $       0.00\pm   0.00$ &   506&  0.00&$  18.4\pm 8.3 $ & 69.2/$ -3.8$/3C410 \\
    8.04& $   0.103\pm   0.004$ & $    25.0\pm     0.1$ & $    3.54\pm    0.17$ & $      82.37\pm   8.82$ &   274&  0.58&$   7.9\pm 7.6 $ & 69.2/$ -3.8$/3C410 \\
    8.43& $   0.042\pm   0.004$ & $   -23.2\pm     0.2$ & $    3.52\pm    0.37$ & $     205.53\pm  15.95$ &   270&  0.59&$  12.5\pm17.9 $ & 69.2/$ -3.8$/3C410 \\
    0.30& $   0.019\pm   0.005$ & $   -46.3\pm     0.3$ & $    2.53\pm    0.68$ & $      16.11\pm  29.14$ &   139&  0.02&$  11.2\pm32.9 $ & 69.2/$ -3.8$/3C410 \\
    2.66& $   0.162\pm   0.008$ & $     2.1\pm     0.0$ & $    1.45\pm    0.06$ & $      17.85\pm  10.98$ &    45&  0.08&$\mathbf{  -0.7\pm 3.8 }$ & 74.5/$-17.7$/3C433 \\
   24.25& $   0.258\pm   0.007$ & $     3.0\pm     0.0$ & $    4.12\pm    0.09$ & $     106.60\pm   4.40$ &   371&  2.21&$\mathbf{  -0.6\pm 3.3 }$ & 74.5/$-17.7$/3C433 \\
    6.71& $   0.076\pm   0.003$ & $     6.9\pm     0.0$ & $    1.97\pm    0.10$ & $      92.22\pm   5.03$ &    84&  0.27&$\mathbf{   6.7\pm 6.6 }$ & 74.5/$-17.7$/3C433 \\
    4.39& $   0.053\pm   0.002$ & $    16.0\pm     0.0$ & $    3.11\pm    0.11$ & $      85.00\pm   4.10$ &   210&  0.27&$  14.7\pm10.8 $ & 74.5/$-17.7$/3C433 \\
    5.91& $   0.093\pm   0.001$ & $     3.9\pm     0.0$ & $    3.12\pm    0.05$ & $      66.22\pm   2.48$ &   213&  0.38&$  -0.3\pm21.2 $ & 88.1/$-35.9$/3C454.0 \\
    6.81& $   0.045\pm   0.001$ & $    -1.5\pm     0.1$ & $    6.39\pm    0.20$ & $     156.34\pm   5.71$ &   892&  0.87&$ -23.1\pm62.3 $ & 88.1/$-35.9$/3C454.0 \\
    7.59& $   0.091\pm   0.001$ & $    -2.0\pm     0.0$ & $    3.69\pm    0.07$ & $      86.97\pm   5.18$ &   297&  0.57&$\mathbf{   6.6\pm 3.6 }$ & 86.0/$-38.1$/3C454.3 \\
    3.50& $   0.079\pm   0.003$ & $     0.7\pm     0.0$ & $    1.81\pm    0.06$ & $      46.32\pm   4.91$ &    71&  0.13&$\mathbf{   3.4\pm 2.9 }$ & 86.0/$-38.1$/3C454.3 \\
    0.88& $   0.022\pm   0.001$ & $     3.4\pm     0.2$ & $    4.20\pm    0.35$ & $      40.89\pm  12.03$ &   386&  0.07&$   5.6\pm15.5 $ & 86.0/$-38.1$/3C454.3 \\
   10.73& $   0.298\pm   0.001$ & $   -10.1\pm     0.0$ & $    2.65\pm    0.01$ & $      41.65\pm   1.52$ &   153&  0.64&$\mathbf{  -2.1\pm 0.9 }$ & 86.0/$-38.1$/3C454.3 \\
    0.42& $   0.048\pm   0.001$ & $   -30.4\pm     0.0$ & $    2.00\pm    0.05$ & $       8.92\pm   3.30$ &    87&  0.02&$\mathbf{   0.8\pm 4.4 }$ & 86.0/$-38.1$/3C454.3 \\
    0.91& $   0.016\pm   0.001$ & $   -35.4\pm     0.1$ & $    3.36\pm    0.18$ & $      56.79\pm   7.80$ &   246&  0.06&$  -1.7\pm16.5 $ & 86.0/$-38.1$/3C454.3 \\
    2.18& $   0.015\pm   0.001$ & $   -16.8\pm     0.1$ & $    5.51\pm    0.27$ & $     151.25\pm   9.11$ &   664&  0.24&$ -51.7\pm23.7 $ & 86.0/$-38.1$/3C454.3 \\
   22.55& $   0.290\pm   0.007$ & $   -10.8\pm     0.1$ & $    5.10\pm    0.13$ & $      89.70\pm   4.14$ &   569&  2.58&$  30.3\pm12.4 $ &157.8/$-48.2$/3C64 \\
    3.71& $   0.086\pm   0.007$ & $     0.2\pm     0.2$ & $    4.37\pm    0.38$ & $      45.23\pm   5.28$ &   417&  0.33&$ -16.7\pm34.8 $ &157.8/$-48.2$/3C64 \\
   17.93& $   0.682\pm   0.006$ & $   -10.4\pm     0.0$ & $    2.32\pm    0.02$ & $      36.26\pm   4.06$ &   117&  1.12&$\mathbf{   5.0\pm 1.5 }$ &170.3/$-44.9$/3C75 \\
    3.33& $   0.095\pm   0.004$ & $    -5.8\pm     0.1$ & $    2.74\pm    0.12$ & $      36.89\pm   3.43$ &   163&  0.19&$\mathbf{ -16.2\pm 8.8 }$ &170.3/$-44.9$/3C75 \\
   11.02& $   0.113\pm   0.003$ & $     5.3\pm     0.1$ & $    5.19\pm    0.13$ & $     103.57\pm   6.97$ &   589&  1.18&$  -1.2\pm10.2 $ &170.3/$-44.9$/3C75 \\
   28.24& $   1.108\pm   0.004$ & $     6.8\pm     0.0$ & $    2.24\pm    0.01$ & $      42.17\pm   7.17$ &   109&  2.04&$\mathbf{  -4.6\pm 1.0 }$ &174.9/$-44.5$/3C78 \\
    4.29& $   0.082\pm   0.001$ & $    10.7\pm     0.1$ & $    4.26\pm    0.14$ & $      54.38\pm   6.31$ &   396&  0.37&$  19.9\pm11.7 $ &174.9/$-44.5$/3C78 \\
    6.81& $   0.153\pm   0.002$ & $     4.2\pm     0.0$ & $    1.92\pm    0.05$ & $      47.89\pm   4.61$ &    80&  0.27&$   2.8\pm 4.5 $ &174.9/$-44.5$/3C78 \\
    6.81& $   0.116\pm   0.002$ & $    -7.7\pm     0.0$ & $    3.07\pm    0.06$ & $      62.07\pm   5.14$ &   206&  0.43&$ -14.3\pm 6.6 $ &174.9/$-44.5$/3C78 \\
    7.09& $   0.297\pm   0.017$ & $     9.5\pm     0.0$ & $    1.35\pm    0.09$ & $      27.59\pm  20.46$ &    39&  0.21&$\mathbf{  -2.8\pm 4.8 }$ &179.8/$-31.0$/3C98 \\
   38.72& $   0.508\pm   0.012$ & $     9.5\pm     0.0$ & $    5.21\pm    0.07$ & $      97.27\pm   6.30$ &   594&  5.02&$\mathbf{  -2.4\pm 3.8 }$ &179.8/$-31.0$/3C98 \\
    5.33& $   0.041\pm   0.003$ & $    23.0\pm     0.2$ & $    6.82\pm    0.58$ & $     134.11\pm   3.95$ &  1016&  0.72&$   4.2\pm41.4 $ &179.8/$-31.0$/3C98 \\
    1.48& $   0.092\pm   0.004$ & $    -1.0\pm     0.1$ & $    4.87\pm    0.22$ & $      16.82\pm   5.14$ &   519&  0.15&$  14.7\pm15.8 $ &179.8/$-31.0$/3C98 \\
   14.24& $   0.600\pm   0.039$ & $     1.4\pm     0.1$ & $    2.04\pm    0.08$ & $      31.55\pm   2.71$ &    90&  0.75&$\mathbf{   5.2\pm 3.6 }$ & 39.6/$ 17.1$/4C13.65 \\
   10.20& $   0.344\pm   0.023$ & $     3.4\pm     0.2$ & $    2.57\pm    0.20$ & $      35.09\pm   3.09$ &   144&  0.60&$\mathbf{  -8.8\pm 6.5 }$ & 39.6/$ 17.1$/4C13.65 \\
   17.73& $   1.161\pm   0.015$ & $     2.0\pm     0.0$ & $    2.10\pm    0.02$ & $      25.81\pm   6.76$ &    96&  1.22&$\mathbf{   4.7\pm 2.3 }$ & 43.5/$  9.2$/4C13.67 \\
   18.90& $   1.019\pm   0.010$ & $     6.2\pm     0.0$ & $    4.01\pm    0.04$ & $      29.56\pm   8.37$ &   351&  2.36&$\mathbf{   5.8\pm 3.3 }$ & 43.5/$  9.2$/4C13.67 \\
    2.36& $   0.030\pm   0.003$ & $    20.2\pm     0.4$ & $    8.20\pm    0.89$ & $      80.73\pm   7.14$ &  1469&  0.38&$  77.6\pm92.0 $ & 43.5/$  9.2$/4C13.67 \\
   67.48& $   4.067\pm   2.174$ & $     7.5\pm     0.1$ & $    0.90\pm    0.15$ & $      68.66\pm  26.77$ &    17&  4.90&$   2.5\pm10.9 $ &188.1/$  0.0$/4C22.12 \\
   60.36& $   7.911\pm   8.309$ & $     4.4\pm     0.1$ & $    0.83\pm    0.19$ & $      60.38\pm  27.61$ &    14&  7.68&$   8.7\pm11.3 $ &188.1/$  0.0$/4C22.12 \\
   35.27& $   6.994\pm   1.782$ & $    -2.2\pm     0.0$ & $    1.26\pm    0.08$ & $      35.31\pm  16.92$ &    34&  6.05&$   7.0\pm 4.5 $ &188.1/$  0.0$/4C22.12 \\
   19.49& $   0.840\pm   0.034$ & $    15.8\pm     0.1$ & $    3.97\pm    0.16$ & $      34.28\pm  12.06$ &   345&  2.23&$  25.3\pm11.0 $ &188.1/$  0.0$/4C22.12 \\
   68.21& $   1.756\pm   0.045$ & $     5.1\pm     0.1$ & $   11.63\pm    0.21$ & $      82.46\pm  27.02$ &  2954& 32.79&$ -55.2\pm13.5 $ &188.1/$  0.0$/4C22.12 \\
   17.87& $   0.166\pm   0.003$ & $     9.6\pm     0.0$ & $    1.87\pm    0.04$ & $     117.02\pm  10.60$ &    76&  0.71&$  -3.1\pm 4.4 $ &186.8/$ -7.1$/P0531+19 \\
   17.77& $   0.474\pm   0.003$ & $     1.8\pm     0.0$ & $    2.09\pm    0.02$ & $      47.10\pm  16.58$ &    95&  0.91&$   0.3\pm 1.9 $ &186.8/$ -7.1$/P0531+19 \\
   39.01& $   0.229\pm   0.001$ & $     5.5\pm     0.0$ & $    8.74\pm    0.06$ & $     190.74\pm  13.06$ &  1670&  7.43&$  15.6\pm 7.1 $ &186.8/$ -7.1$/P0531+19 \\
    2.81& $   0.073\pm   0.001$ & $    -3.6\pm     0.0$ & $    2.23\pm    0.02$ & $      40.00\pm   1.10$ &   108&  0.13&$ -29.5\pm19.8 $ &222.5/$ 63.1$/P1055+20 \\
   16.34& $   1.285\pm   0.042$ & $    33.9\pm     0.0$ & $    1.70\pm    0.05$ & $      22.59\pm  10.30$ &    63&  0.96&$  -0.8\pm 2.1 $ &201.5/$  0.5$/T0629+10 \\
   14.47& $   0.271\pm   0.016$ & $    30.9\pm     0.1$ & $    2.82\pm    0.26$ & $      60.90\pm  13.83$ &   173&  0.91&$  31.2\pm 8.8 $ &201.5/$  0.5$/T0629+10 \\
   13.92& $   0.354\pm   0.017$ & $    23.3\pm     0.1$ & $    3.03\pm    0.16$ & $      46.68\pm  19.93$ &   201&  0.98&$  -2.8\pm 7.1 $ &201.5/$  0.5$/T0629+10 \\
   28.44& $   1.605\pm   0.035$ & $     4.9\pm     0.0$ & $    4.37\pm    0.06$ & $      35.60\pm  14.16$ &   416&  4.86&$ -16.9\pm 3.0 $ &201.5/$  0.5$/T0629+10 \\
   46.16& $   0.297\pm   0.005$ & $    16.9\pm     0.3$ & $   28.09\pm    0.46$ & $     179.87\pm   8.78$ & 17249& 29.19&$  45.1\pm24.0 $ &201.5/$  0.5$/T0629+10 \\
    3.96& $   0.152\pm   0.017$ & $   -11.9\pm     0.1$ & $    1.29\pm    0.16$ & $      28.13\pm   7.36$ &    36&  0.11&$   4.0\pm 7.7 $ &201.5/$  0.5$/T0629+10 \\
   71.60& $   0.346\pm   0.003$ & $     5.2\pm     0.0$ & $   10.91\pm    0.06$ & $     245.06\pm   4.27$ &  2602& 18.00&$\mathbf{   2.1\pm 3.3 }$ &184.6/$ -5.8$/TauA \\
    8.37& $   1.242\pm   0.007$ & $    10.7\pm     0.0$ & $    2.09\pm    0.01$ & $      11.77\pm   1.06$ &    95&  0.59&$\mathbf{  -3.1\pm 0.6 }$ &184.6/$ -5.8$/TauA \\
    2.38& $   0.355\pm   0.008$ & $     5.1\pm     0.0$ & $    1.79\pm    0.04$ & $       7.98\pm   4.20$ &    69&  0.10&$\mathbf{   7.2\pm 1.7 }$ &184.6/$ -5.8$/TauA \\
   18.01& $   0.507\pm   0.005$ & $     2.8\pm     0.0$ & $    2.61\pm    0.05$ & $      45.33\pm   2.88$ &   148&  1.17&$\mathbf{   2.1\pm 1.4 }$ &184.6/$ -5.8$/TauA \\
    0.00& $   2.634\pm   0.105$ & $   -48.0\pm     0.0$ & $    4.22\pm    0.08$ & $       0.00\pm   0.00$ &   388&  0.00&$\mathbf{   9.3\pm 0.5 }$ &111.8/$ -2.1$/CasA \\
    0.00& $   1.598\pm   0.040$ & $   -38.0\pm     0.0$ & $    6.67\pm    0.11$ & $       0.00\pm   0.00$ &   971&  0.00&$\mathbf{  25.0\pm 0.8 }$ &111.8/$ -2.1$/CasA \\
   16.86& $   1.299\pm   0.047$ & $    -0.7\pm     0.0$ & $    2.89\pm    0.07$ & $      23.19\pm   1.63$ &   183&  1.70&$\mathbf{  -0.3\pm 0.6 }$ &111.8/$ -2.1$/CasA \\[4pt]
\enddata
    
\tablecomments{The ordering is by source name as follows: 3C, 4C, P, T,
TauA, CasA. Temperatures are in K, velocities in km s$^{-1}$,
column densities in $10^{20}$ cm$^{-2}$, magnetic fields in $\mu$G. }
    
\end{deluxetable}


\begin{deluxetable}{ccccccccc} 
\tablewidth{0pc} 
\tablecaption{Dependence of $B_{||}$ on Cloud Ordering \label{ordering} }
\tablehead{ 
\colhead{SOURCE/VLSR} & 
\colhead{$B_{||,0}$} & 
\colhead{$B_{||,max}$} & 
\colhead{$B_{||,min}$}
}

\startdata TauA/10.7   &  $-3.09 \pm 0.55$  & $-3.66 \pm 0.65 $ & $
-3.09 \pm 0.55 $ \\  TauA/5.1    &  $ 7.19 \pm 1.69$  & $10.56 \pm 2.50
$ & $ 7.10  \pm 1.67 $ \\  3C409/15.7  &  $ 5.74 \pm 1.06$  & $8.03 \pm
1.47 $ & $ 5.74  \pm 1.01 $ \\  CasA/--48.0 &  $ 9.29 \pm 0.54$  & $9.74
\pm 0.55 $ & $ 8.70  \pm 0.55 $ \\  \enddata \tablecomments{ $B_{||,0}$
is from Table \ref{mkzmntable_mod}; $B_{||,max}$ and $B_{||,min}$ are
the largest and smallest fields derived including opacity effects (see
\S \ref{yetanother}). Magnetic fields are in $\mu$G and VLSR in km s$^{-1}$.}
\end{deluxetable}

\end{document}